\documentclass{emulateapj}

\shorttitle{A Ly$\alpha$+He\textsc{ii} Nebula at $\lowercase{z}\approx1.67$}
\shortauthors{Prescott, Dey, \& Jannuzi}

\newcommand{\lya}{Ly$\alpha$}
\newcommand{\heii}{He~\textsc{ii}}
\newcommand{\civ}{C~\textsc{iv}}
\newcommand{\ciii}{C~\textsc{iii}]}
\newcommand{\siiv}{Si~\textsc{iv}}
\newcommand{\nv}{N~\textsc{v}]}
\newcommand{\neiv}{Ne~\textsc{iv}]}

\newcommand{\popii}{Pop~II}
\newcommand{\popiii}{Pop~III}

\newcommand{\lae}{LAE}

\newcommand{\bw}{$B_{W}$}

\begin{document}

\title{The Discovery of a Large L\lowercase{y}$\alpha$+H\lowercase{e}~II Nebula at $\lowercase{z}\approx1.67$: A Candidate Low Metallicity Region?}

\author{Moire K. M. Prescott\altaffilmark{1}, Arjun Dey\altaffilmark{2}, and Buell T. Jannuzi\altaffilmark{2}}

\altaffiltext{1}{Steward Observatory, University of Arizona, Tucson, AZ 85721; mprescott@as.arizona.edu}
\altaffiltext{2}{National Optical Astronomy Observatory, 950 North Cherry Avenue, Tucson, AZ 85719; dey@noao.edu, jannuzi@noao.edu}

\begin{abstract}

We have discovered a $\approx$45~kpc \lya\ nebula (or \lya\ ``blob'') at $z\approx1.67$ which 
exhibits strong, spatially-extended \heii\ emission and very weak 
\civ\ and \ciii\ emission.  This is the first spatially-extended \lya+\heii\ emitter observed and the 
lowest redshift \lya\ blob yet found.  
Strong \lya\ and \heii$\lambda$1640 emission in the absence of metal lines has been 
proposed as a unique observational signature of primordial galaxy formation (e.g., from gravitational cooling 
radiation or Population~III star formation), but no convincing examples of spatially-extended \lya+\heii\ 
emitters have surfaced either in \lya-emitting galaxy surveys at high redshifts ($z>4$) or in studies of 
\lya\ nebulae at lower redshifts.  From comparisons with photoionization models, we find that the observed line 
ratios in this nebula are consistent with low metallicity gas ($Z\lesssim10^{-2}-10^{-3} Z_{\sun}$), 
but that this conclusion depends on the unknown ionization parameter of the system.  
The large \heii\ equivalent width ($\approx$37$\pm$10\AA) and the large \heii/\lya\ ratio (0.12$\pm$0.04) suggest 
that the cloud is being illuminated by a hard ionizing continuum, either an AGN or very low metallicity stars, 
or perhaps powered by gravitational cooling radiation.  
Thus far there is no obvious sign of a powerful AGN in or near the system, so in order 
to power the nebula while remaining hidden from view even in the mid-infrared, 
the AGN would need to be heavily obscured.  
Despite the strong \lya+\heii\ emission, it is not yet clear what is the dominant power source for this nebula.  
The system therefore serves as an instructive example of how the complexities of true astrophysical 
sources will complicate matters when attempting to use a strong \lya+\heii\ signature as a unique 
tracer of primordial galaxy formation.
\end{abstract}


\section{Introduction}

Understanding primordial galaxy formation is a major science driver for the next generation of large 
space- and ground-based telescopes and has inspired a substantial amount of theoretical literature due to the 
potential contribution of the first generations of stars to reionization and the early stages of galaxy evolution.  
According to theoretical predictions, the observational signpost of primordial galaxy formation 
is the presence of strong \lya$\lambda$1216 and \heii$\lambda$1640 emission lines, 
either due to photoionization by very low metallicity and 
Population~III (\popiii, i.e., zero metallicity) star formation \citep[e.g.,][]{tum01,sch03,sch08} 
or due to gas cooling during gravitational collapse \citep[e.g.,][]{hai00,yang06}.  
In the case of \popiii\ stars, the strong \lya\ and \heii\ is a direct consequence of the low metallicities, 
where, in the absence of metals, H and He become 
the dominant line coolants for the gas, and of the hot effective temperatures of \popiii\ stellar clusters, which are 
predicted to show a top-heavy IMF and low stellar atmospheric opacity due to the lack of metals \citep[e.g., ][]{ez71,bro01}.  
In the case of gravitational cooling radiation, \lya\ and \heii\ are the primary ways for pristine gas to cool 
as it is collisionally excited during gravitational collapse; the predicted \heii/\lya\ ratios may 
be as high as 10\% \citep[e.g., ][]{hai00,yang06}.  
Strong \lya\ and \heii\ emission lines are commonly seen in other astrophysical sources with hard 
ionizing continua, but at normal metallicities they are generally accompanied by strong metal lines 
such as \ciii\ and \civ, as seen for example, in AGN, radio galaxy halos, Wolf-Rayet galaxies, or cases 
of shock ionization \citep[e.g., ][]{reu07,lei96,dop96}.  
The presence of strong \lya\ and \heii\ emission in the absence of strong metal lines 
has been put forward as a potentially unique observational signature of primordial galaxy formation.

Although theoretical studies suggest that \popiii\ and very low metallicity star formation 
may persist down to lower redshifts, this depends on the feedback efficiency, i.e., 
the ability of a \popiii\ stellar population to pollute the 
large-scale surroundings with metals \citep[e.g.,][]{tor07}.  
Searches for \popiii\ stars have understandably 
pushed to higher redshift ($z>4$), where the \popiii\ star formation rate density should increase 
dramatically relative to that found in the local Universe \citep{sca03}.  
Thus far, no unambiguous case of a \popiii\ stellar population has been observed.  
Several \lya-emitting galaxy studies at $z\sim4-5$ have uncovered sources with unusually 
high \lya\ equivalent widths ($W_{rest}>240$ \AA) -- larger than that expected from a normal stellar population -- 
suggesting either a top-heavy IMF, a very low metallicity, and/or a very young age \citep{mal02,rho03}.    
However, the complicated radiative transfer of \lya\ in a clumpy ISM could also be responsible for boosting 
the \lya\ \citep{fink08}, and in these studies, no corroborating evidence for the \popiii\ scenario in the form of 
a strong \heii\ detection was found in either the individual or stacked \lae\ spectra, 
leaving the matter unresolved \citep{daw04,ouch08}.  Deep spectroscopic observations of a strong 
\lya-emitting galaxy at $z\approx6.33$ showed no evidence for \heii\ emission \citep{nag05}, and a 
more recent \lya+\heii\ dual emitter survey at $z\sim4-5$ found no convincing candidates \citep{nag08}.  
The limits from each of these studies suggest that \popiii\ star formation does not dominate in these $z\sim4-5$ 
samples, prompting the authors to encourage searches at ever higher redshifts ($z\gtrsim7$).  Additional 
high redshift \lya+\heii\ surveys are underway \citep[e.g., ][]{ali08}.

In this paper we report on the discovery of a $z\approx1.67$ \lya\ nebula with strong, spatially-extended 
\heii\ emission and very weak \civ\ and \ciii\ emission.  
This is the first spatially-extended source that resembles the predicted \lya+\heii\ signature 
of primordial galaxy formation.  
However, the system is more complex than it first appears.  
The observed line ratios suggest that the 
nebula may contain low metallicity gas, but this depends on the unknown 
ionization parameter of the system.  
Detailed analysis of the spectra along with 
extensive multi-wavelength data reveals that the source of ionization is uncertain:  
the nebula is either an H~\textsc{ii} region 
ionized by a hard spectrum source, i.e., an AGN or a very low metallicity stellar population, 
or a gravitationally cooling cloud.  
The fact that multiwavelength follow-up observations are required in order to 
better constrain the source(s) of ionization and metallicity of the nebula has implications 
for \lya+\heii\ searches at higher redshift.  
In Section 2, we summarize the systematic search for \lya\ nebulae that led to this 
discovery and our observations and reductions. 
Section 3 contains a discussion of the observational results, and Section 4 details our analysis 
of the physical properties of the nebula.  
In Section 5 we discuss the implications of this discovery for the ongoing high redshift 
\lya+\heii\ surveys, and we summarize our conclusions in Section 6.

We assume the standard $\Lambda$CDM cosmology ($\Omega_{M}$=0.3, $\Omega_{\Lambda}$=0.7, $h$=0.7);
1\arcsec\ corresponds to a physical scale of 8.47~kpc at $z=$~1.671.  All magnitudes are in the AB system.  
Unless otherwise stated, \heii\ refers to \heii$\lambda$1640\AA, \civ\ to \civ$\lambda\lambda$1549,1550, 
\ciii\ to \ciii$\lambda$1909, and \neiv\ to \neiv$\lambda$2424.

\section{Observations \& Reductions}

In this section we discuss the observations that led to the discovery of this \lya+\heii\ nebula 
and the methods used to process the imaging and spectroscopic data.  

\subsection{The Search}

Large \lya\ nebulae (or \lya\ ``blobs") $-$ large ($\sim$100~kpc) clouds of gas emitting strongly in \lya\ $-$ 
are thought to be sites of ongoing galaxy formation and have been found in small numbers, primarily at $2<z<3$.  
Early theoretical work suggested that these nebulae could be examples of gravitationally cooling clouds \citep{hai00}, 
and more recent cosmological hydrodynamic simulations indicated that cooling clouds should be detectable as 
\lya+\heii\ nebulae, although the specific predictions proved uncertain due to the treatment of 
star-forming gas (Yang et al. 2006; Y. Yang 2008, private communication).  
Several of the largest \lya\ nebulae, including one with strong \heii\ and \civ\ emission \citep{dey05}, 
have since been shown to be powered instead by AGN, spatially-extended star formation, 
or some combination \citep[e.g., ][]{dey05,mat07,gea07}.  
Two groups claim to have discovered \lya\ nebulae that are powered by gravitational cooling 
radiation \citep{nil06,smi07,smi08}, but neither case shows strong \heii\ emission.  

\lya\ nebulae are extremely rare objects and have often been found using deep narrow-band imaging surveys of 
known galaxy overdensities.  As such, their space density, particularly at the bright end of the luminosity function, 
is largely unconstrained.  In order to carry out an efficient but unbiased survey of a large cosmic volume, we have 
designed a systematic morphological search for spatially-extended \lya\ nebulae using the {\it broad-band} imaging 
from the NOAO Deep Wide-Field Survey \citep[NDWFS; ][]{jan99} Bo\"otes Field.  This field has been imaged in 
\bw, $R$, and $I$-band with median 5$\sigma$ point-source depths of $\approx$27.1, 26.3, and 25.8 mag (AB), 
respectively.  In the sub-field relevant to this paper, the \bw, $R$, and $I$-band 5$\sigma$ point-source depths 
are 27.5, 26.1, and 26.0 mag for 2.3, 1.7, and 2.8~hrs of integration, respectively.  Candidates were selected 
from the broad-band data using a morphological and color selection algorithm, and spectroscopic follow-up was 
used to rule out low-redshift interlopers.  The advantage of using broad-band data to search for line-emitting 
sources, a seemingly crude approach, is the enormous comoving volumes ($\sim$10$^{8}$~h$_{70}^{-3}$~Mpc$^{3}$) 
that can be surveyed efficiently using publicly available data over wide fields.  The success of the present 
survey relied heavily on the depth of the broad-band NDWFS imaging and the darkness of the sky within 
the \bw-band, against which strong line emission can dominate the flux even within the very broad \bw\ filter.  

A full discussion of the search algorithm and results 
will be discussed elsewhere (Prescott et al. 2009, in preparation).  
Here we report on the discovery of a new \lya+\heii\ nebula at $z\approx1.67$ (hereafter denoted PRG1) located 
in the Bo\"otes field at 14:35:12.439 +35:11:07.16 (J2000).  
Unlike the other \lya\ sources found in our systematic search, the \lya+\heii\ nebula presented here 
was selected by the search algorithm not because of its \lya\ emission, which lies outside the \bw\ filter, but 
instead because of the strong and diffuse blue continuum emission ($\approx$92\%) and 
spatially-extended \heii\ emission ($\approx$8\%) within the bandpass.  
Postage stamps from GALEX \citep[FUV \& NUV;][]{mar05}, the NOAO Deep Wide-Field Survey (\bw, $R$, and $I$), 
and the Spitzer Deep Wide-Field Survey (SDWFS; IRAC 3.6$\mu$m, 4.5$\mu$m, 5.8$\mu$m, and 8.0$\mu$m; 
Ashby et al. 2009, submitted) are shown in Figure~\ref{fig:prg1postage}.  The GALEX limits are 0.36~$\mu$Jy in the NUV and FUV bands; 
the SDWFS limits are 3.2, 4.4, 25.5, and 25~$\mu$Jy (5$\sigma$).  The MIPS coverage of this region shows 
no detection with a 1$\sigma$ rms limit of 51~$\mu$Jy (E. Le Floc'h 2008, private communication).

\subsection{Spectroscopic Follow-up}

\begin{figure}
\plotone{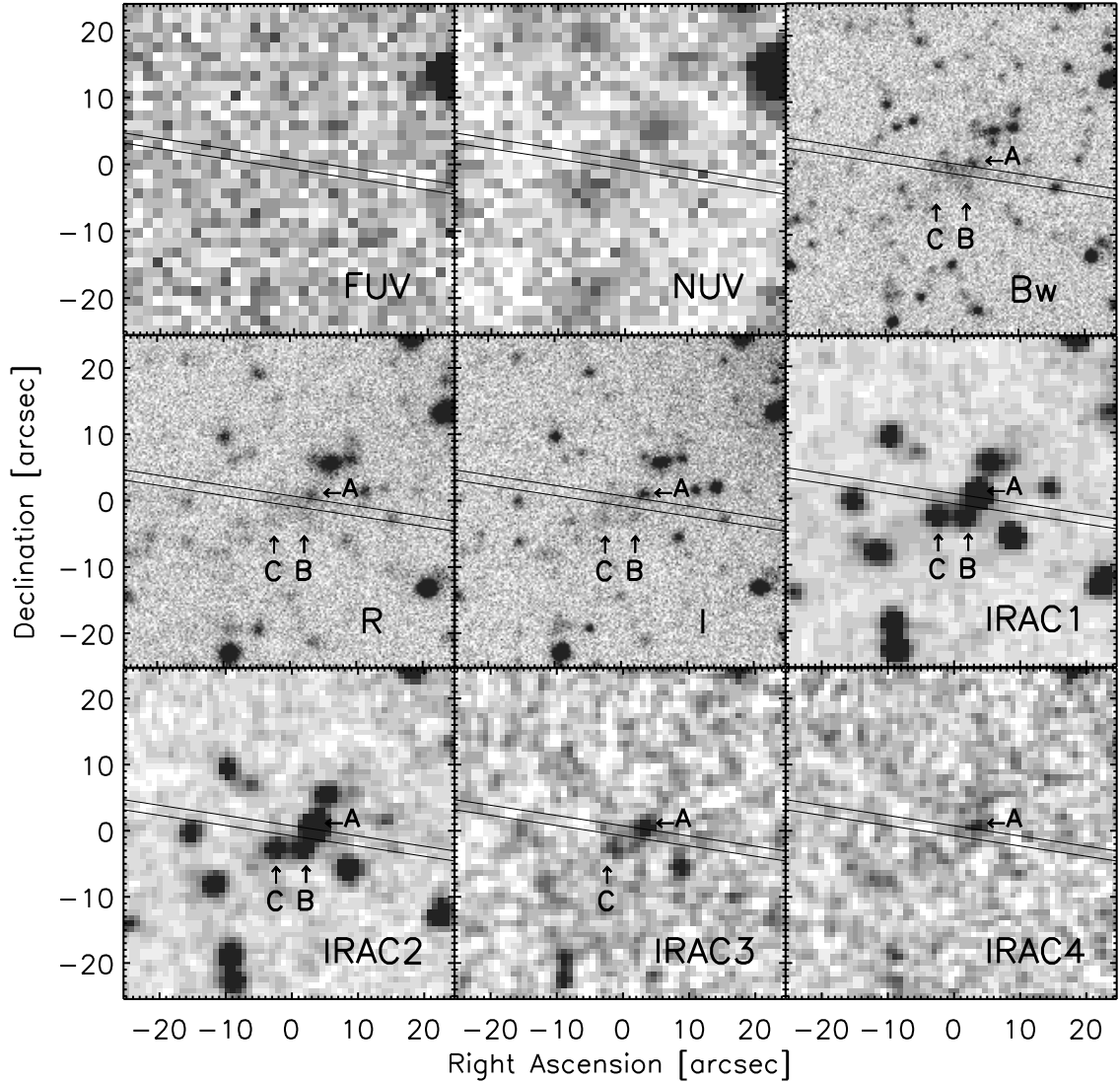}
\caption{GALEX ($FUV$ and $NUV$), NDWFS broad-band optical 
(\bw, $R$, and $I$), and IRAC (3.6$\mu$m, 4.5$\mu$m, 5.8$\mu$m, 
and 8.0$\mu$m) postage stamps for PRG1.  Images are all 1\arcmin\ on a side 
and centered on the coordinate location 14:35:12.439 +35:11:07.16 (J2000).  
The spectroscopic slit is shown with a position angle of 81.2$^{\circ}$.  
PRG1 was selected as a \lya\ nebula candidate due to the diffuse blue emission in the \bw\ imaging.  
However, in this case \lya\ is in fact outside the \bw\ band; the \bw\ flux is instead dominated by diffuse continuum 
and \heii\ emission.  The diffuse blue continuum emission is clearly visible in the NDWFS \bw-band imaging.  
Several compact sources in the near vicinity of the nebula have IRAC counterparts, labeled A, B, and C.}
\label{fig:prg1postage}
\end{figure}

We obtained spectroscopic follow-up observations using the MMT and the Blue Channel Spectrograph 
during UT 2008 June 8-9 as part of our systematic search for \lya\ nebulae.  We used a 1.5$\times$120\arcsec\ 
(unvignetted) slit and the 300 l/mm grating ($\lambda_{c}\approx5713$\AA, $\Delta\lambda\approx3100-8320$\AA).  
We chose a slit orientation that spanned the longest dimension of the diffuse emission (PA=81.2$^{\circ}$, 
observed near transit), as shown in Figure~\ref{fig:prg1postage}, while also intersecting a nearby bright 
object.  We moved the target along the slit by $\approx$5\arcsec\ inbetween exposures.

Conditions during the first night 
were clear and stable with 1\farcs0 seeing; the second night was clear but with variable seeing ($\gtrsim$1\farcs3) 
and high winds which caused shaking of the telescope pointing.  We show in Section 3.2 that due 
to a slight pointing offset and wind-shake, the data from the second night sample a 
different spatial region within the nebula 
and are contaminated at some level by sources nominally off the slit, in particular a red compact 
source to the WNW (Source A in Figure~\ref{fig:prg1postage}).  For this reason, data from the two nights were 
reduced and analyzed separately.  The most robust \lya\ and \heii\ flux measurements are from the first night, 
when conditions were excellent during the single half hour exposure.  An additional 1.5 hours of integration were 
obtained during the second night.  The \lya\ flux measurement from Night 2 shows a 25\% loss relative to that 
from Night 1.  In the remainder of the paper we use the \lya\ and \heii\ flux measurements from Night 1 but 
include the line ratios derived from both nights.   

The data were reduced using IRAF\footnote{IRAF is distributed by 
the National Optical Astronomy Observatories, which are operated by the Association of Universities for 
Research in Astronomy, Inc., under cooperative agreement with the National Science Foundation.}.  
After performing the overscan and bias subtraction, we corrected flatfield exposures for the response 
of the internal `Bright Continuum' flatfield lamp by dividing out the median along columns and 
then applied the flatfield correction. 
We used twilight flats to determine the illumination correction for the science frames.  
Cosmic rays were removed using {\it xzap}\footnote{http://iraf.noao.edu/iraf/ftp/iraf/extern/xdimsum020627}.  
The wavelength solution was determined using HeArNe and HgCd comparison lamps, with an rms of $\approx$0.17\AA.  
We corrected the data for a slight systematic offset in the night sky lines; the night sky line wavelengths in 
the final spectra are correct to $\pm$0.3\AA.  Flux calibration was based on observations of the standard stars 
BD+33 2642 and Wolf~1346\footnote{KPNO IRS Standard Star Manual}.  
We applied a grey shift ($\lesssim$0.08~mag) and fit the sensitivity function using extra 
care at the blue end of the spectrum because the \lya\ line at $\approx$3250\AA\ lies only 56.3~ pixels (109.7\AA) 
from the edge of the chip.  
The instrumental resolution measured from the Hg~\textsc{i}$\lambda$4047 line is 3.6\AA, 
and the tilt within the aperture is $\le\pm$5.24~km s$^{-1}$ over the region of the nebula.  

\section{Results}
\subsection{Ly$\alpha$ and He~II Emission}
\begin{figure}
\plotone{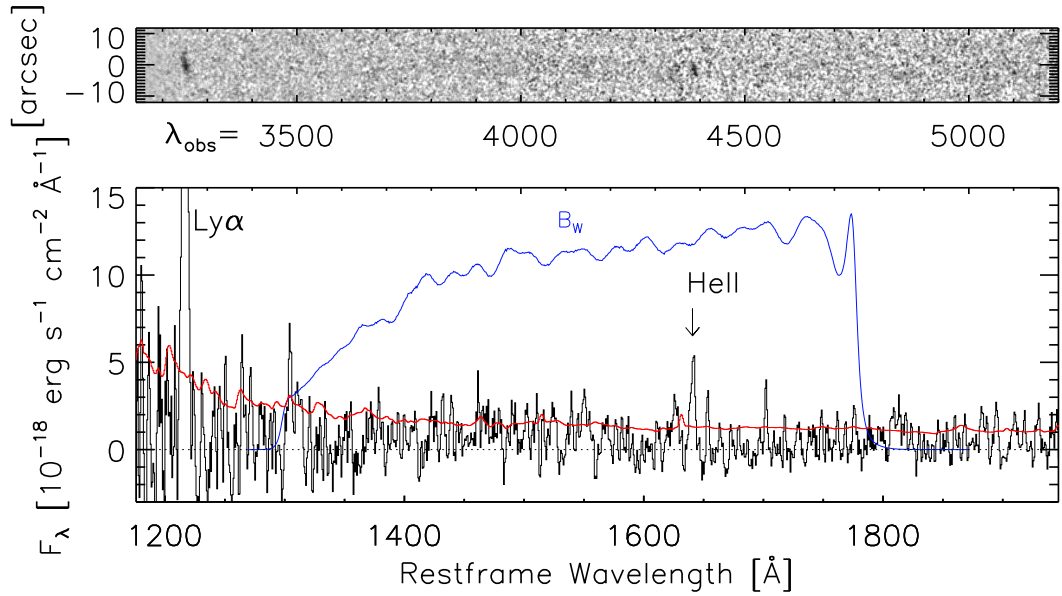}
\caption{Night~1 (UT 2008 June 08) 2D and 1D spectra showing strong \lya\ and \heii\ emission from a single half hour exposure (1.5$\times$5.0\arcsec\ aperture).  
The spectra have been smoothed by 0.84\arcsec\ spatially and by 5.8 \AA\ in the spectra dimension.  
The filter curve is the \bw\ bandpass convolved with the atmospheric transmission (blue).  
The 1$\sigma$ error spectrum is overlaid (red).}
\label{fig:night1spec}
\end{figure}

\begin{figure}
\plotone{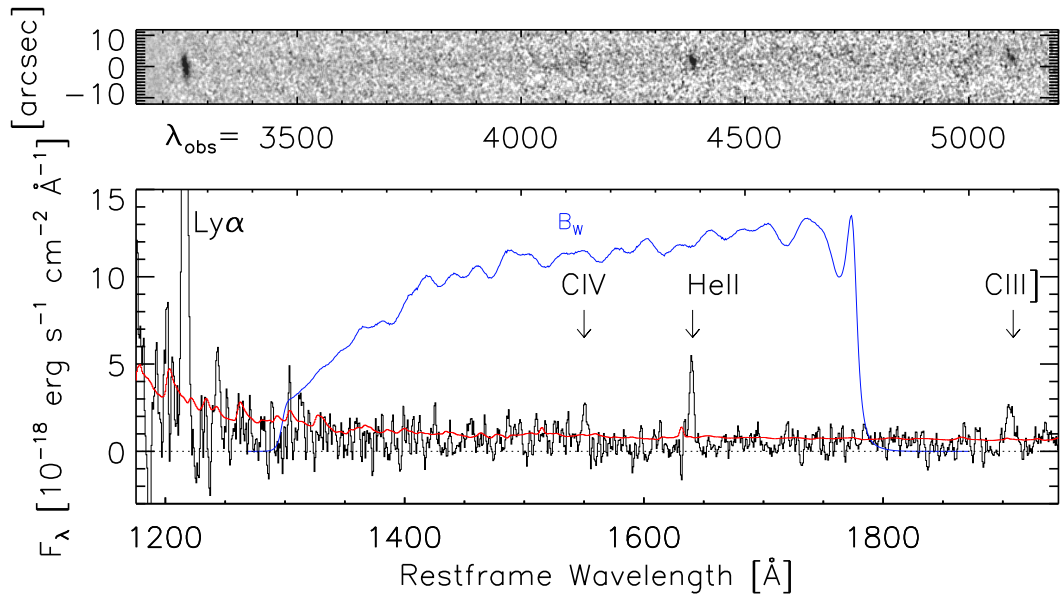}
\caption{Night~2 (UT 2008 June 09) 2D and 1D spectra showing strong \lya\ and \heii\ emission, weak \ciii, and 
marginally-detectable \civ\ from a combined 1.5 hour exposure (1.5$\times$5.0\arcsec\ aperture).  
The spectra are smoothed by 0.84\arcsec\ spatially and by 5.8 \AA\ in the spectral dimension.  
The filter curve is the \bw\ bandpass convolved with the atmospheric transmission (blue).  
The 1$\sigma$ error spectrum is overlaid (red).  When comparing to Figure~\ref{fig:night1spec}, note that a 180$^{\circ}$ change 
in the slit position angle between the two nights caused a flip in the angle of the spectral trace across the detector.}
\label{fig:night2spec}
\end{figure}

The final 2D and 1D spectra from the first and second night are shown in 
Figures~\ref{fig:night1spec}-\ref{fig:night2spec}.  The spectra show strong \lya\ and \heii\ 
emission lines, both of which exhibit similar kinematic structure in the 2D spectrum.  
The measured fluxes and flux limits were derived separately from each night's data using 
a 1.5$\times$5.0\arcsec\ aperture, chosen to maximize the signal-to-noise ratio of the 
\heii\ measurement (Table~\ref{tab:fluxkin}).  
Faint continuum emission is detected in the spectra (Figure~\ref{fig:night2spec}).  
We scaled the spectra by a factor of 1.4 and 1.6, respectively, in order to match the 
continuum fluxes measured from the much deeper NDWFS broad-band data 
within the region covered by the slit (Table~\ref{tab:phot}).  

The \lya\ and \heii\ luminosities from Night 1 are $L_{Ly\alpha}=L_{Ly\alpha,ap}\times f_{geo}\times 
f_{profile}\approx5.4\times10^{43}$~erg~s$^{-1}$ and 
$L_{He\textsc{ii}}=L_{He\textsc{ii},ap}\times f_{geo}\approx4.0\times10^{42}$~erg~s$^{-1}$.   
$L_{Ly\alpha,ap}=9.3\times10^{42}$~erg~s$^{-1}$ 
and $L_{He\textsc{ii},ap}=1.2\times10^{42}$~erg~s$^{-1}$ are the luminosities measured 
within the spectroscopic aperture, $f_{geo}$ is the geometric correction factor between the spectroscopic aperture and 
full extent of the nebula, and $f_{profile}$ corrects for asymmetry in the \lya\ profile due to blue side absorption.  
We discuss the estimation of these correction factors below.  

\begin{figure}
\plotone{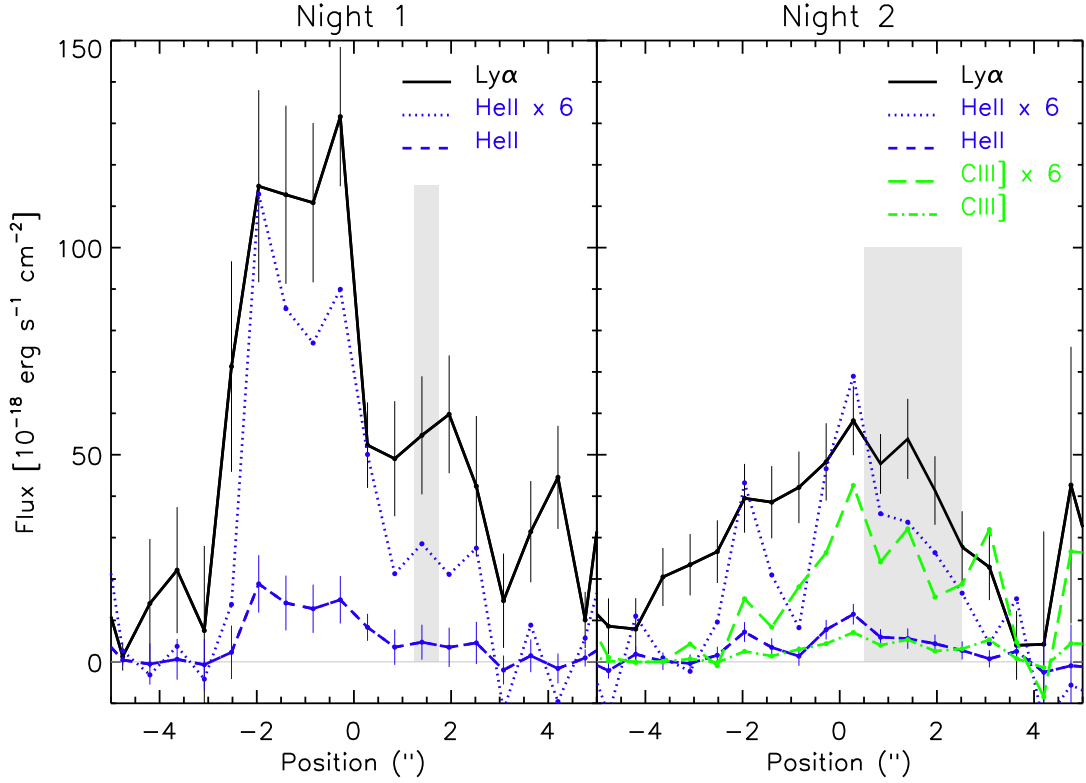}
\caption{\lya\ and \heii\ spatial profiles with errors as a function of position along the slit 
from the Night 1 (UT 2008 June 08) and Night 2 (UT 2008 June 09).  As \ciii\ was not detected on Night 1, the \ciii\ profile is shown for 
Night 2 only.  \heii\ and \ciii\ profiles scaled by a factor of 6 are also overplotted to allow easier comparison 
with the \lya\ profile.  The \heii\ region appears to be comparable in size to the \lya\ region.  
The shaded area represents the approximate position of Source A, a source 
that was off-slit but that may have contaminated the Night 2 observations due to poor seeing and 
windy conditions.  The discrepancy between the spatial profiles and the offset of the \ciii\ spatial profile 
relative to the \lya\ are both consistent with the hypothesis that the Night 2 spectrum is contaminated 
by light from Source A and that Source A may be the primary source of the metal line emission.}
\label{fig:spatialprofile}
\end{figure}

The geometric correction factor $f_{geo}$ was derived from a comparison of the spatial 
extent of the nebula measured in the \lya\ line versus the extent in the \bw\ imaging.  
In our best seeing Night 1 data, where we have accurate spatial information and low slit losses, 
the \lya\ is extended by $\approx$5\arcsec\ (42.3 kpc) with a fairly sharp 
truncation at large radii, but with a possible extension towards the west (right of center, Figure~\ref{fig:spatialprofile}).  
The \heii\ emission is weaker than the \lya\ by a factor of 10, but it appears that 
at lower signal-to-noise the \heii\ nebula extends to roughly the same radius as the 
core of the \lya\ nebula (no westward extension).  The Night 2 data show a \lya\ extent of $\approx$6\arcsec\ 
and a \heii\ extent of $\approx$5\arcsec\ but suffer from slit losses and 
degraded spatial resolution due to poor seeing and wind-shake.    
The extent of the \lya\ nebula in the spectroscopic observations is in rough agreement with the 
spatial extent of the diffuse emission measured from the broad-band \bw\ data along the 
position of the slit ($\approx$6.6\arcsec, $\approx$56~kpc above a \bw\ surface brightness of 
4.5$\times10^{-16}$ erg s$^{-1}$ cm$^{-2}$ arcsec$^{-2}$).  
The full area of the diffuse emission in the \bw\ imaging is 
$\approx26$ square arcseconds.  Assuming that the \lya\ emission is distributed similarly to the \bw\ continuum emission, 
we estimated that a geometric correction factor of $f_{geo}=3.4$ is required to obtain the total 
\lya\ flux from the nebula.  
This correction is very approximate; narrow-band imaging 
and/or spatially-resolved spectroscopy will be required to accurately account for 
the contributions of line and continuum emission to different portions of the nebula.  

\begin{figure}
\plotone{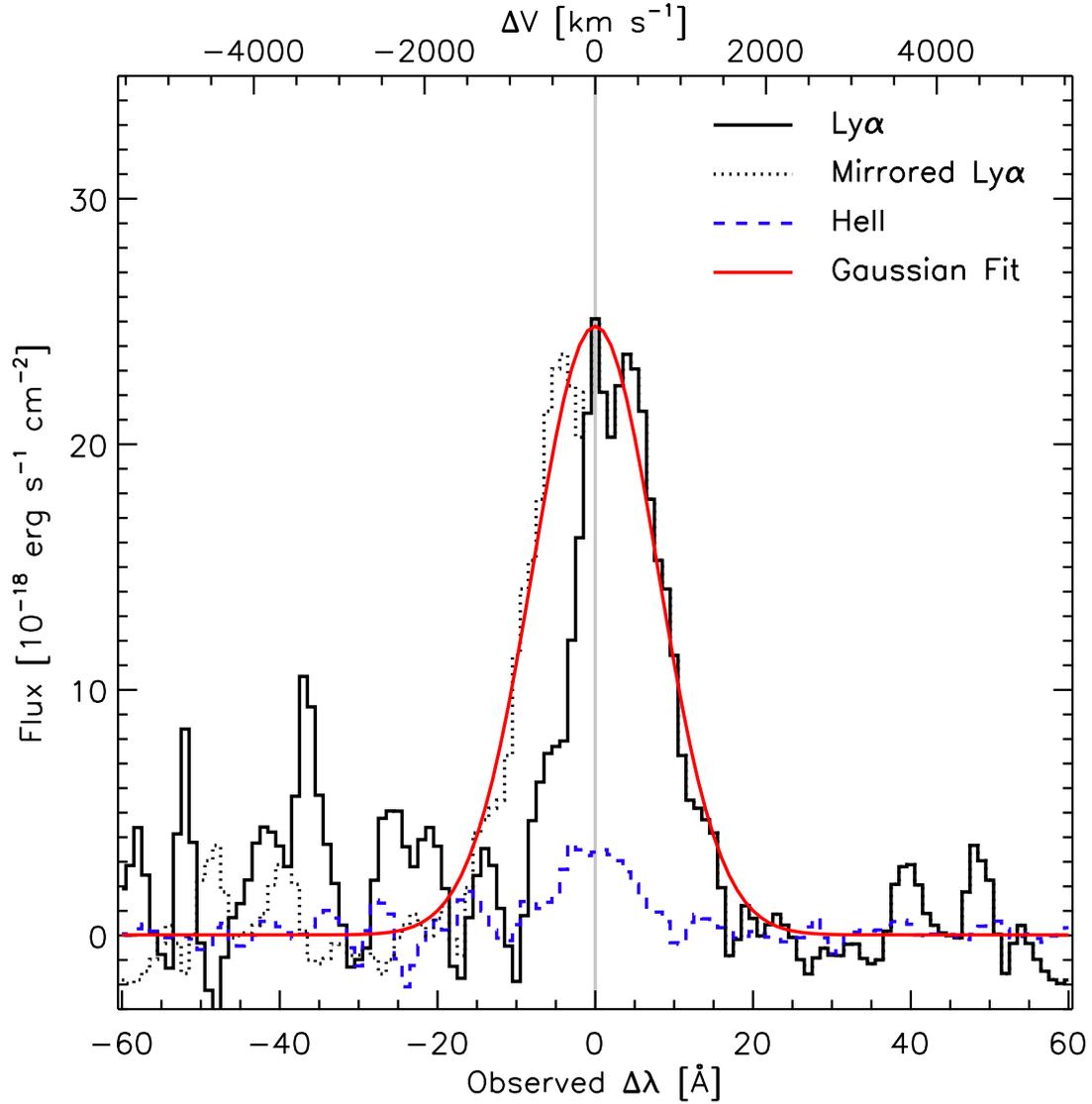}
\caption{\lya\ (black solid line) and \heii\ (blue dashed line) line profiles as a function of observed wavelength 
centered on the systemic redshift of the system, as measured from the \heii\ line.  
The observed \lya\ is shown with the mirror image of the 
long-wavelength half of the line superimposed (black dotted line).  
A Gaussian fit (red solid line) indicates that the \lya\ line may be 
absorbed by $\sim$41\%.  The corresponding velocity offsets for \lya\ are given on the top axis.  }
\label{fig:lyaheiiprofile}
\end{figure}

We derived the \lya\ blue side absorption correction $f_{profile}$ using a comparison 
of the \lya\ and \heii\ line centroids.  The \lya\ profile is fairly symmetric (Figure~\ref{fig:lyaheiiprofile}), but 
the center of the \lya\ line is offset to the red from the systemic redshift, as determined 
from the centroid of the  \heii\ emission line, likely due to absorption.  
We estimated the amount of blue side absorption of \lya\ by mirroring the red side 
of the \lya\ profile across the line centroid, and derived 
a factor of 1.7 (Night 1) and 1.8 (Night 2) increase in the \lya\ flux.  We therefore used a 
correction factor of $f_{profile}=1.7$ to obtain the final Night 1 \lya\ luminosity.  

\begin{figure}
\plotone{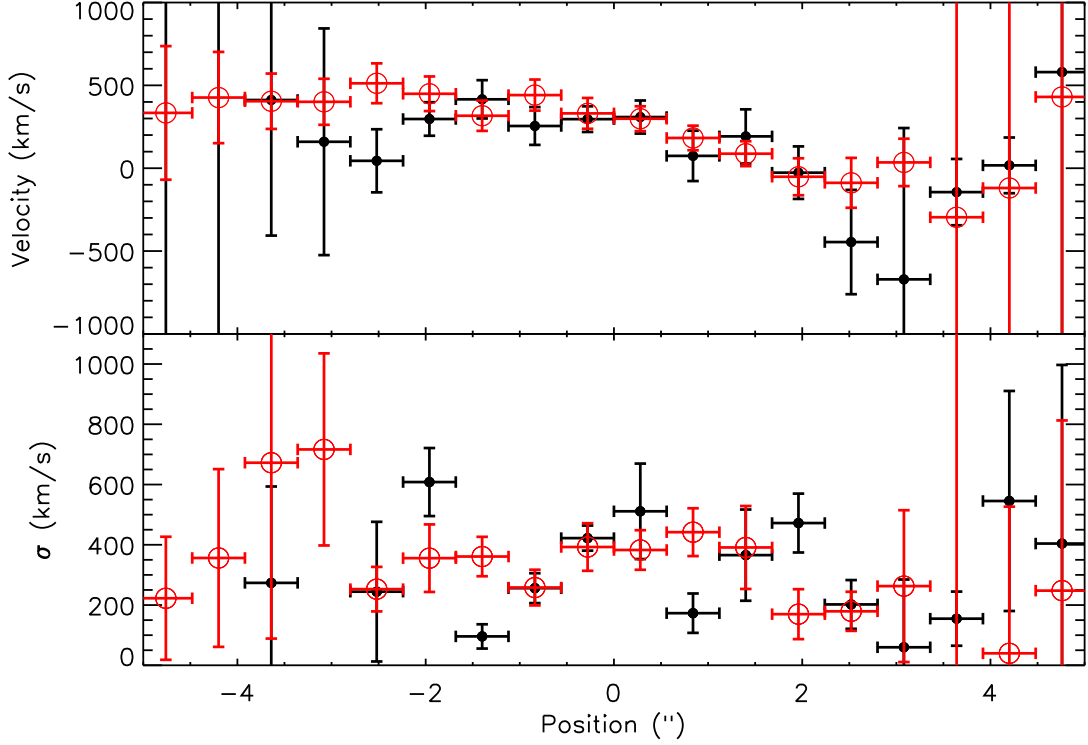}
\caption{Velocity and velocity dispersion spatial profiles of the \lya\ line for Night 1 (UT 2008 June 08; black filled circles) 
and Night 2 (UT 2008 June 09; red open circles) derived from spectra extractions taken in 2~pixel (0.56\arcsec) 
spatial bins and corrected for the instrumental resolution.  The velocity zeropoint was set using 
the redshift calculated from the \heii\ line centroid in the full 5.0\arcsec\ aperture extraction.}
\label{fig:veloprofile}
\end{figure}

The nebula shows clear velocity structure in both the \lya\ and \heii\ lines.  
The lines are resolved, showing \lya\ and \heii\ velocity dispersions of 
$\sigma_{v}\approx400$ and 250~km s$^{-1}$, 
respectively, corrected for the instrumental resolution.  
Figure~\ref{fig:veloprofile} shows the velocity profile of the \lya\ line from Night 1 
and Night 2 derived using 2~pixel (0.56\arcsec) extractions.  The profile is relatively smooth but 
flattens on the east (left) of center.  The spatially-resolved \lya\ velocity dispersion 
is essentially constant across the nebula.

\subsection{C~IV, C~III], and Ne~IV] Emission}
Despite the strong \lya\ and \heii\ emission seen in the Night 1 data, there is no detection of 
\ciii, \civ, or \neiv.  The 1$\sigma$ upper limits on the line ratios 
are \civ/\heii$<0.23$, \ciii/\heii$<0.19$, and \neiv/\heii$<0.23$.  
Due to the excellent and stable observing conditions during Night 1, the Night 1 
spectrum provides the most accurate flux measurements for \lya\ and \heii\ along with higher resolution 
kinematic information and the strongest limits on the \ciii, \civ, and \neiv\ emission from the source.  
In contrast, the Night 2 spectrum shows weak \ciii\ and marginal \civ\ and \neiv\ emission at 
the same redshift.  The resulting line ratios are \civ/\heii~$=0.36$, \ciii/\heii~$<0.82$, and \neiv/\heii~$<0.49$.   
The Night 2 data provide additional constraints on the line ratios but must be treated with care 
due to the Night 2 observing conditions.  Due to the poor seeing and the wind-shake of the telescope, 
the Night 2 spectrum suffers from slit losses as well as contamination 
from nearby sources nominally off the slit, most importantly from a compact red source to the WNW of the 
target center (Source A; see Figure~\ref{fig:prg1postage}).  

From a comparison of the line ratios derived from each night, we argue that the 
metal line emission is not from the same spatial location as 
the \lya\ and \heii, and may instead be associated with the region closer to Source A.  
On Night 2, the \ciii/\heii\ ratio was 0.82; thus if the emission were perfectly cospatial, 
we should have detected \ciii\ on Night 1 at $\approx$5.1$\times10^{-17}$ erg s$^{-1}$ cm$^{-2}$, 
roughly the same significance as \heii.  Instead, we can rule out \ciii\ emission at the 4.3$\sigma$ level.  
For \neiv, we should have detected it at 3.0$\times10^{-17}$ erg s$^{-1}$ cm$^{-2}$ 
on Night 1, which disagrees at the 2.2$\sigma$ level with our Night 1 result.  
We cannot make a robust comparison for \civ, as it is only detected at the 2$\sigma$ level even on Night 2, 
but we make the assumption that all the metal line emission originates from the same source.  
In contrast, the \heii/\lya\ line ratios are 
consistent between the two nights (0.12$\pm$0.04 and 0.13$\pm$0.02, respectively), indicating that the \lya\ 
and \heii\ are indeed cospatial within the region sampled by these observations even 
though the metal line emission varies spatially.  

The offsets in the spatial profiles of the lines are consistent with the idea that 
the data from the two nights sampled different spatial regions.  
If the \civ, \ciii, and \neiv\ lines are from the region around 
Source A and if the wind-shake was consistently perpendicular to the 
slit, there should be a $\approx$5.2~pixel (1.5\arcsec\ West) offset between the spatial centroid of these lines and 
that of \lya.  However, this offset will vary by an estimated $\approx\pm$3.6~pixels ($\approx$1\arcsec) or more depending 
on the direction of the telescope wind-shake relative to the angle of the slit during a given exposure.  
We do see a difference between the two nights when we look at the spatial profiles extracted in 2~pixel spatial 
bins along the spatial direction (Figure~\ref{fig:spatialprofile}); the Night 1 profile is skewed towards the 
East (left, away from Source A), whereas the Night 2 profile is peaked closer to the position 
of Source A, suggesting contamination.  We use these spatial profiles to 
compute flux-weighted mean spatial centroids for each emission line separately.  
The \heii\ position is consistent with that of \lya\ to within 
1~$\sigma$ ($\Delta$x=1.41$\pm$1.54~pixels, 0.40$\pm$0.43\arcsec).  
The \ciii\ offset is 4.03$\pm$2.07~pixels (1.13$\pm$0.58\arcsec) in the direction of Source A.  
(Due to the intrinsic faintness of the \civ\ and \neiv\ lines, the computed offsets are 
not statistically significant.)  Follow-up observations will be required to resolve this issue, but it appears 
that the \ciii\ line is offset from the spatial centroid of the \lya\ at the 1.9$\sigma$ level, 
in the right direction and at roughly the correct position to be explained by contamination from 
the region closer to Source A.  

Given the pointing uncertainty and inevitablity of contamination from sources off the slit due to the unstable conditions 
on Night 2, the discrepancy in the spatial profiles between the two nights, 
the evidence that the \ciii\ line shows a spatial offset consistent with the position of 
Source A, and the fact that the Night 1 data are inconsistent with the fluxes of \neiv\ and \ciii\ measured 
on Night 2 at $\sim$2-4$\sigma$, we argue that much of the metal line emission 
derives from the region around Source A.  We cannot rule out that some is emitted further out in the 
nebula, but even if it does, it arises from a region spatially distinct from the region 
observed on Night 1.  Combining the data from the two nights would not 
be appropriate due to the different spatial sampling of the observations and 
these intrinsic spatial inhomogeneities.  In the absence of more definitive data, we 
use the metal emission line measurements from both nights as independent upper limits on the 
emission coming from two possibly distinct regions of the nebula.

\section{Discussion}

In this section we derive estimates for the physical properties of PRG1 
and use photoionization models to gain insight into the gas metallicity 
and the possible source(s) of ionization.  
Despite the strong \lya+\heii\ signature and weak metal line emission often 
associated with primordial phenomena, we find 
that it is not possible to make an unambiguous determination of the source of ionization.  
The line ratios are consistent with a nebula comprised of low  
(but non-zero) metallicity gas, irrespective of the nature of the ionizing source, 
but this conclusion depends on the unknown ionization parameter of the system.

\subsection{Physical Properties of the Nebula}

Our discovery data can be used to put constraints on the physical properties of the nebula.  
Using the emission measure of \lya, we can estimate the electron density: 
\begin{eqnarray}
L_{Ly\alpha}&=\frac{j_{Ly\alpha}}{j_{H\beta}}n_{p}n_{e}fVh\nu_{H\beta}\alpha^{eff}_{H\beta} \\
&\approx 1.2 \frac{j_{Ly\alpha}}{j_{H\beta}}n_{e}^{2}fVh\nu_{H\beta}\alpha^{eff}_{H\beta}
\end{eqnarray} 
where $j_{Ly\alpha}$ and $j_{H\beta}$ are the emission coefficients for \lya\ and H$\beta$, respectively, 
$n_{p}$ and $n_{e}$ are the proton and electron number densities with $n_{e}\approx 1.2 n_{p}$ (the factor 
of 1.2 accounts for the contribution of electrons from doubly-ionized Helium), $f$ is the 
volume filling factor of the nebula, $V$ is the volume of the nebula, $h$ is Planck's constant, $\nu_{H\beta}$ 
is the frequency of $H\beta$, and $\alpha^{eff}_{H\beta}$ is the effective recombination coefficient for $H\beta$ \citep{ost89}.  
We approximate the nebula as a sphere with radius $R\approx28.0$~kpc.  
The \lya\ luminosity extrapolated to the entire nebula and corrected for 
blue side absorption ($5.4\times10^{43}$~erg~s$^{-1}$) corresponds to an electron number density of 
$n_{e}\approx 0.094 f^{-0.5}$~cm$^{-3}\approx 29.7 (f/10^{-5})^{-0.5}$~cm$^{-3}$, where we have 
used a typical value for $f$ derived for the line-emitting regions in cluster cooling flows 
\citep[$\sim10^{-5}$; e.g., ][]{heck89}.  
This corresponds to an ionized gas mass of 
$M_{ion}$=1.25$m_{p}n_{e}fV=$8.4$\times10^{8}(f/10^{-5})^{0.5}~M_{\sun}$.  

Similarly, the \heii\ emission measure can be used to estimate the He$^{++}$ and electron densities, 
assuming the cosmic mass fraction of He:
\begin{eqnarray}
L_{He\textsc{ii}\lambda1640}&=\frac{j_{\lambda1640}}{j_{\lambda4686}}n_{He^{++}}n_{e}fVh\nu_{\lambda4686}\alpha^{eff}_{\lambda4686} \\
&\approx\frac{j_{\lambda1640}}{j_{\lambda4686}} (14) n_{He^{++}}^2 fVh\nu_{\lambda4686}\alpha^{eff}_{\lambda4686} 
\end{eqnarray} 
where $j_{\lambda1640}$ and $j_{\lambda4686}$ are the emission coefficients, 
$n_{He^{++}}$ and $n_{e}$ are the He$^{++}$ and electron number densities with $n_{e}\approx 1.2 n_{p} \approx 14 n_{He^{++}}$, 
$\nu_{\lambda4686}$ is the frequency of \heii$\lambda4686$, and $\alpha^{eff}_{\lambda4686}$ 
is the \heii$\lambda4686$ effective recombination coefficient \citep{ost89}.  
The \heii\ luminosity extrapolated to the whole nebula  ($4.0\times10^{42}$~erg~s$^{-1}$) corresponds to 
a He$^{++}$ number density of $n_{He^{++}}\approx 1.0 (f/10^{-5})^{-0.5}$~cm$^{-3}$ and 
an electron number density of $n_{e}\approx 14.5 (f/10^{-5})^{-0.5}$~cm$^{-3}$.
This corresponds to an ionized gas mass of 
$M_{ion}$=1.25$m_{p}n_{e}fV=$ 4.1$\times10^{8}(f/10^{-5})^{0.5}$ M$_{\sun}$, roughly consistent 
with the \lya\ estimate.  

If we make the naive assumption that the velocity spread results from rotation 
with $V_{c}\approx\Delta V/2=350$~km~s$^{-1}$ at a radius of 28~kpc (3.3\arcsec), we estimate 
the mass interior to this radius to be $M_{rot}=$8.0$\times$10$^{11}$sin$^{2}i~M_{\sun}$, 
where $i$ is the inclination of the system.  
If the velocity dispersion is due to random motions of small clouds within the system, we estimate the dynamical mass 
(using the more kinematically robust \heii\ line; $\sigma_{v}=245$ km s$^{-1}$) 
to be $M_{rand}=1.9\times10^{12}$ M$_{\sun}$.  

The spatial extent of the \lya\ (H$^{+}$) and \heii\ (He$^{++}$) emitting regions appear 
to be similar in size, certainly within a factor of two 
($R_{H^{+}}/R_{He^{++}}\lesssim2$).  This is somewhat surprising when we consider a simple Str\"omgren 
sphere argument.  Assuming a hard central ionizing source (an AGN or a \popiii\ SED) embedded within an 
infinite spherical H+He cloud, the predicted size ratio of the H$^{+}$ and He$^{++}$-emitting regions 
is $R_{H^{+}}/R_{He^{++}}\sim9-17$, 
an order of magnitude higher than we observe.  This suggests that either the H$^{+}$ region is density-bounded, that the sources 
of ionization are distributed throughout the nebula, or that the assumption of spherical symmetry is 
invalid (e.g., the cloud is illuminated from the outside).  

The observed \lya\ luminosity ($L_{Ly\alpha}=5.4\times$10$^{43}$ erg s$^{-1}$) 
corresponds to a H-ionizing photon flux ($13.6$~eV~$\leq$~E$_{\gamma}\leq54.4$~eV) of: 
\begin{eqnarray}
Q(H)=\frac{L_{Ly\alpha}}{h\nu_{Ly\alpha}}\frac{1}{0.68}\approx4.9\times10^{54}~photons~s^{-1}
\end{eqnarray}
where we have assumed that the fraction of H ionizing photons converted into \lya\ is 0.68 \citep{spi78}.  
We note that this is likely a lower limit due to a number of considerations: 
\lya\ is highly susceptible to resonant scattering and is easily destroyed by dust, and \lya\ for 
this system is observed at $\approx$3250\AA, a wavelength regime that suffers from very low atmospheric 
transimission and poor CCD sensitivity, making accurate flux calibration difficult.  

From the observed \heii\ emission ($L_{He\textsc{ii}}=4.0\times$10$^{42}$ erg s$^{-1}$), 
we calculate a He$^{+}$-ionizing photon flux (E$_{\gamma}\geq54.4$~eV) of: 
\begin{eqnarray}
Q(He^{+})=\frac{L_{\lambda1640}}{h\nu_{\lambda1640}}\frac{\alpha^{eff}_{He\textsc{ii}}}{\alpha^{1640}_{He\textsc{ii}}}\approx6.2\times10^{53}~photons~s^{-1}
\end{eqnarray}
where $\alpha^{eff}_{He\textsc{ii}}=1.53\times10^{-12}$ cm$^{3}$ s$^{-1}$ 
\citep[case B; 100~cm$^{-3}$, 10$^{4}$~K;][]{stor95} and 
$\alpha^{\lambda1640}_{He\textsc{ii}}=\alpha^{4686}_{He\textsc{ii}}
\frac{j_{\lambda1640}}{j_{\lambda4686}}\frac{\nu_{\lambda4686}}{\nu_{\lambda1640}}=8.08\times10^{-13}$ 
cm$^{3}$ s$^{-1}$ \citep[case B;][]{ost89}.  

The large value of Q(He$^{+}$)/Q(H)=0.13 is strong evidence that the 
source is illuminated by a hard ionizing continuum.   
The prediction for a \popii\ stellar population \citep[instantaneous burst, Salpeter IMF, 1-100 M$_{\Sun}$, 
$Z=0.001$;][]{sch03} is Q(He$^{+}$)/Q(H)$=0.0004$; while our \lya\ measurement may be compromised by 
radiative transfer effects and poor flux calibration at the edge of the atmospheric cut-off, 
it would need to have been underestimated by more than a factor of 100 in order for the observed 
Q(He$^{+}$)/Q(H) to match that of a normal \popii\ stellar population.  
The observed Q(He$^{+}$)/Q(H) is in the range expected for AGN or \popiii\ stars.  A typical 
AGN template has a ratio of Q(He$^{+}$)/Q(H)$\sim$0.09 \citep{elv94}, more consistent with 
the observational constraint.  
While normal metallicity stellar populations will have very little flux above 
the He$^{+}$ ionization edge, \popiii\ stars are predicted to have much harder spectra 
due to high temperatures, low stellar atmospheric opacity, and a top-heavy IMF.  
Strong mass loss from these stars would cause higher effective tempertures, boosting 
the hard ionizing photons even further.  
\citet{sch02} calculated a suite of \popiii\ models, both with and without strong mass loss.  
For individual high mass stars (M$\geq$80-300~M$_{\sun}$) they predict Q(He$^{+}$)/Q(H)$\geq$0.022-0.11.  
Models with mass loss yields ratios of Q(He$^{+}$)/Q(H)$\geq$0.06-0.17 (80-300~M$_{\sun}$).  
However, such large Q(He$^{+}$)/Q(H) ratios persists for only a few Myrs for instantaneous burst models.  
Constant star-forming \popiii\ models (with no mass loss) integrated over a range of IMFs yield 
Q(He$^{+}$)/Q(H)$\leq$0.04, and the value decreases with increasing metallicity \citep{sch03}.  
There are numerous uncertainties in these estimates, but broadly speaking 
the observed Q(He$^{+}$)/Q(H) ratio is in the range populated only by AGN and the 
very lowest metallicity stellar populations.  

\subsection{Photoionization Modeling}

The ubiquity of AGN, the association of AGN and extended emission line regions (EELRs), 
and the presence of spatially varying metal line emission 
suggest that an AGN is a plausible source of ionization.  
At the same time, the observed line ratios are highly unusual and have been tied in the theoretical 
literature to primordial galaxy formation processes (the presence of \popiii\ stars or gravitational 
cooling radiation).  A comparison of the line ratios from PRG1 with those of radio galaxy EELRs 
and other \lya\ nebulae is shown in Table~\ref{tab:lineratio}.  While the \lya/\heii\ ratios 
are comparable, other than the \ciii/\heii\ ratio on Night 2, all the metal line ratios from 
this source are at the low end or lower than the range seen in EELRs.  
Furthermore, this source shows very different line ratios than those seen in 
another large radio-quiet \lya\ nebula at $z\approx2.7$ \citep{dey05}.  

In order to gain greater insight into the metallicity and possible ionization sources for the nebula, 
we used {\tt Cloudy}\footnote{Calculations were performed with version 07.02.02 of {\tt Cloudy}, 
last described by \citet{fer98}.} to model simple, constant density gas clouds being illuminated by AGN, \popiii, 
and \popii\ SEDs and predict the resulting line ratios and continuum emission.  
The AGN template is taken from \citet{mat87} ($F_{\nu}\propto\nu^{-1.0}$ at $26<$~h$\nu<56$ eV, 
$F_{\nu}\propto\nu^{-3.0}$ at $56<$~h$\nu<365$ eV), 
and the \popiii\ spectra are \citet{tum06} 1~Myr models 
($Z=0$, a top-heavy IMF peaked around 10$M_{\sun}$ with $\sigma$=1.0, i.e., their case A).  
The \popii\ case is a 1~Myr, $Z=0.001$, Salpeter IMF, instantaneous burst model from Starburst 99 
\citep{lei96}.  

The strong, spatially-extended \lya+\heii\ emission and weak, spatially-variable \ciii\ and \civ\ emission in our discovery 
spectra and the blue, spatially-extended continuum emission measured from deep broad-band imaging  
provide constraints on the metallicity of the gas and on the slope of the ionizing continuum in the system.  
He$^{+}$ and C$^{+3}$ have similar ionization potentials (54.4~eV and 47.9~eV, respectively), 
so a comparison of \heii\ and \civ\ puts constraints on the metallicity that are less 
dependent on the ionizing continuum slope.  The \lya\ and \ciii\ emission lines 
(relevant ionization potentials of $H$ and $C^{+2}$: 13.6~eV and 24.4~eV) provide additional 
constraints on the slope of the ionizing continuum.  

The observed lines will also depend on the ionization parameter of the system.  
A very rough estimate based on our discovery data is $U = \phi(H)/(n_{H} c) \gtrsim 0.0002$, 
where $\phi$(H)=Q(H)/($\pi R^2$) is the surface flux of ionizing photons, n$_{H}$ is the total hydrogen gas density, 
and $R$ is the radius of the cloud.  We have assumed our observed quantities: 
Q(H)=4.9$\times10^{54}$~photons s$^{-1}$, n$_{H}\sim n_{e}\sim29.7$~cm$^{-3}$, and $R\approx28$~kpc.  
However, this estimate is uncertain and most likely a lower limit.  We have argued that 
the \lya\ flux is likely underestimated and raised the possibility that the 
system is density-bounded, both of which will cause an underestimate of the ionizing photon flux 
(possibly by an order of magnitude).  
In addition, the geometry of the system is clearly complex, and the estimated density depends critically 
on the assumed value of the filling factor.  The density estimates in Section 4.1 are larger than typical values 
for the ISM ($n_{H}\sim$1~cm$^{-3}$); if the cloud is in fact more similar to typical ISM densities, 
the ionization parameter would increase by over an order of magnitude.  
A more sophisticated treatment of the ionization parameter is beyond the scope of this analysis, 
so for the purposes of this paper we modeled a reasonable range of ionization parameters: Log~U=[-3,~-1,~0].

\subsection{Metallicity of the Gas}
\begin{figure}
\plotone{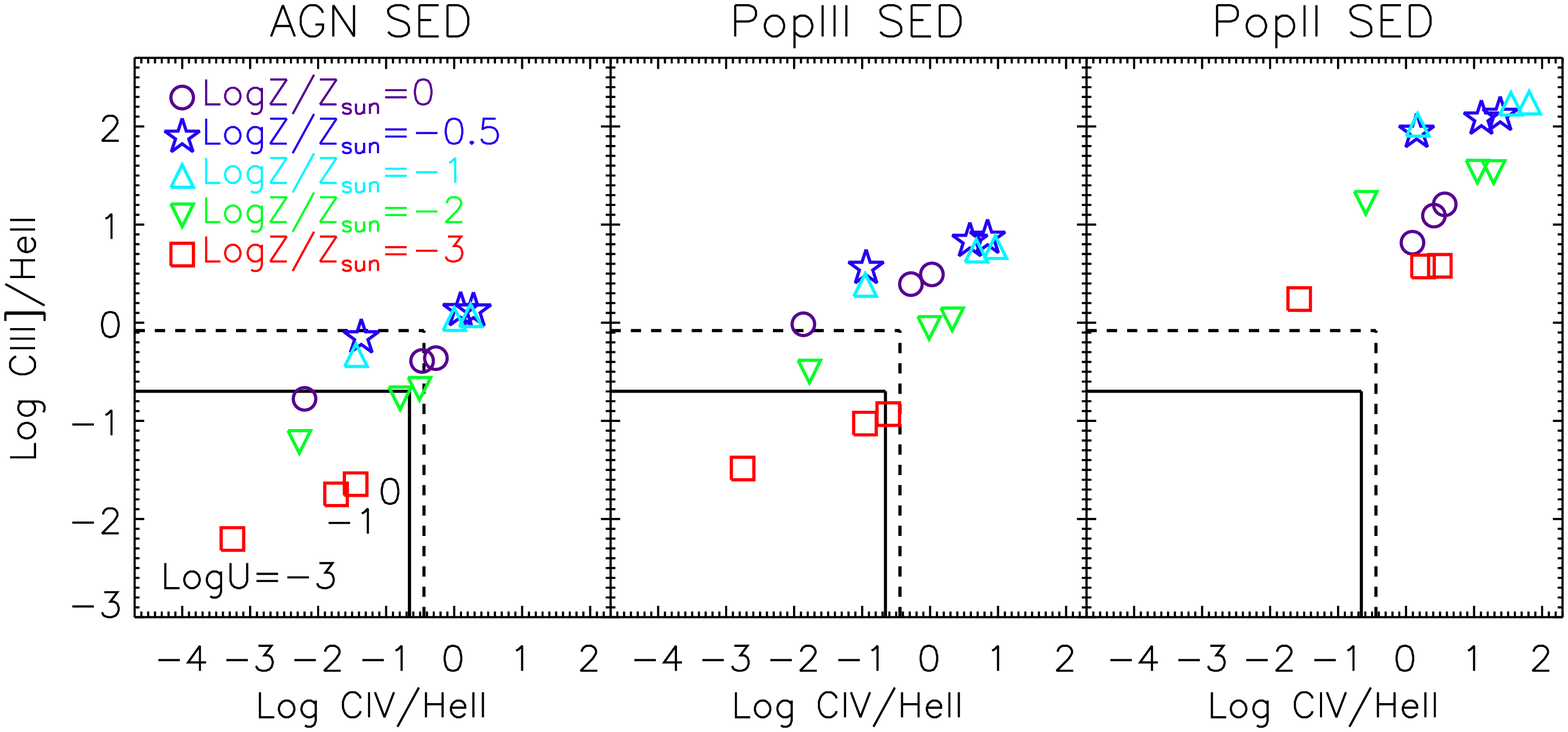}
\caption{\ciii/\heii\ versus \civ/\heii\ line ratios for a grid of {\tt Cloudy} models over a range of SEDs, 
ionization parameters, and gas metallicities.  
The models represent a cloud of gas with $Log Z/Z_{\sun}=$[-3, -2, -1, -0.5, 0] being ionized with an 
ionization parameter of $Log U=$[-3, -1, 0] by one of three SEDs: 
an AGN SED (left), a \citet{tum06} Population~III galaxy model SED (middle; Z=0, 1~Myr burst, moderately 
top-heavy IMF discussed in the text), and a Starburst99 Population~II galaxy model SED 
\citep[right; Z=0.001, Salpeter IMF, 1~Myr burst; ][]{lei96}.  
The 1$\sigma$ limits on the line ratios of the nebulosity are shown for Night~1 (solid line) and Night~2 (dashed line).}
\label{fig:cloudysimple}
\end{figure}

In order to explore the range of parameter space allowed by our observations, 
in Figure~\ref{fig:cloudysimple} we plot line diagnostics for models 
with a range of ionization parameters, metallicities, 
and ionizing SEDs, comparing them to our observed limits on the line ratios 
of the nebulosity from Night 1 and Night 2.  

For low ionization parameters (Log~U~$<-1$), it is possible to produce the observed 
line ratios with an AGN SED illuminating solar metallicity gas (Figure~\ref{fig:cloudysimple}; blue plus signs).  
For higher ionization parameter (Log~U~$\geq-1$), the observed ratios require low metallicity gas.  
In the case of an AGN SED, the \civ/\heii\ and \ciii/\heii\ limits imply $Z<10^{-2} Z_{\sun}$.  
Models with \popiii\ ionization lead to lower metallicity estimates of $Z<10^{-3} Z_{\sun}$.  
(For the \popii\ model even lower metallicities would be required, but this 
case is highly unlikely given the large Q(He$^{+}$)/Q(H) ratio, as discussed in Section 4.1.)  
To put this into context, these metallicity estimates are at or below the 
lowest limits for weak Mg~\textsc{ii} absorbers at $0.4<z<2.4$ 
\citep[$>$10$^{-2}$-10$^{-2.5}$;][]{lyn07}.  
Studies of the most metal-poor damped \lya\ absorbers have shown that none have 
metallicities lower than $[Fe/H]>-3$ \citep{pet08}.  
\lya\ forest clouds at $z\approx3$ with N(H\textsc{i})$>$10$^{15}$~cm$^{-2}$ are 
uniformly metal-enriched with carbon abundances of $\approx10^{-2}~Z_{\sun}$, 
and observations of lower column density \lya\ forest clouds indicate that 
there may be universal minimum metallicity of 3.5$\times$10$^{-4}~Z_{\sun}$ 
that is roughly constant from $z\approx2-6$ \citep{son01}.  
If the metallicity of PRG1 is similar to 
the lowest metallicities measured from absorption-line studies, this would be the first 
time such a system has been seen in \lya\ and \heii\ emission.

While the weak metal line emission suggests a low metallicity system, 
the metallicity estimates are uncertain due to a variety of factors, none of which are 
well-constrained by the current data, e.g., the ionization parameter, 
the geometry of the cloud and ionizing source(s), and the intensity and spectrum of the source(s) of ionization.  
Putting stronger constraints on the metallicity of the nebula will require deeper spectroscopy 
and more detailed photoionization modeling, which is beyond the scope of the current paper.

\subsection{Source of Ionization}

The large \heii\ equivalent width and large \heii/\lya\ ratio 
is strong evidence that the nebula is powered by a hard ionizing continuum.  
We discuss each of the possible ionization sources in turn: AGN, shock ionization, Wolf-Rayet stars, 
low metallicity (\popii) and zero metallicity (\popiii) star formation, and 
gravitational cooling radiation.

\subsubsection{AGN}

An AGN can produce high \heii/\lya\ ratios and weaker 
\civ\ and \ciii\ emission lines, particularly if it is illuminating a low metallicity cloud.  
This scenario is certainly plausible, as emission line halos (e.g., \lya, \ciii, \heii, \civ) around radio galaxies 
have been known for some time, arising predominantly from a combination of jet-ISM interactions and scattered light from the AGN 
or from spatially-extended star formation \citep[e.g., ][]{mcc87,vano96,dey97,vil03,reu03}.  
Unlike our \lya+\heii\ nebula, however, 
these gaseous haloes are predominantly metal-enriched, with strong \ciii\ and \civ\ emission 
\citep[e.g.,][]{ reu07, max02}.
  
While there is no compact source visible in the center of the nebula, there are several 
compact sources in the vicinity of the nebula that could in principle harbor an AGN 
(Figure~\ref{fig:prg1postage}).  For two reasons we believe that if there is an AGN 
in the system, it must be at the position of Source A.  
First, we find that even if all the nearby sources were AGN, Source A would 
contribute the vast majority of the ionizing photon flux.  When we scale 
the standard quasar template from \citet{elv94} to match the 
measured \bw\ flux from each source and calculate the corresponding ionizing photon flux, the net 
contribution is only 2\% of the ionizing photon flux required to explain the \lya, with nearly all of that arising from Source A.  
Second, in Section 3.2 we argued that the metal line emission is not cospatial with the \lya+\heii\ 
nebula, and that the observed metal lines likely derive from the region of Source A.  
Below we discuss the likelihood that an AGN at the position of Source A is 
powering the \lya+\heii\ nebula.  

If we assume that the Night~2 metal line emission stems primarily from Source A, 
we can compare the measured line ratios to those of well-studied galaxy populations.  
At face value, the \ciii\ and \civ\ emission lines associated with the region around Source A are suggestive of an AGN, 
however we find that the line ratios are more consistent with that seen in non-AGN LBGs \citep{shap03}.  
The measured ratios of \civ/\lya=0.05$\pm$0.03 and \ciii/\lya=0.11$\pm$0.02 (uncorrected for \lya\ absorption) 
are likely upper limits due to the uncertainties 
in the \lya\ measurement.  Even so, the \civ/\lya\ ratio is a factor of 4-5 lower than what is seen in 
LBGs flagged as narrow-line AGN \citep[\civ/\lya$\approx$0.25;][]{shap03} and in local Seyfert galaxies 
\citep[\civ/\lya$\approx$0.21;][]{fer86}.  
In addition, if \lya\ is underestimated by a factor of two, the corrected \civ/\lya\ and \ciii/\lya\ ratios 
would match those of non-AGN LBGs.  Furthermore, the ratio of \ciii/\civ=2.3$\pm$1.2 
is in agreement with that found for non-AGN LBGs \citep[\ciii/\civ$\approx$2.5;][]{shap03} and is a factor of four higher than expected from an 
narrow-line AGN spectrum \citep[\ciii/\civ$\approx$0.05;][]{shap03}, suggesting Source A has 
a softer ionizing continuum.

There is currently no evidence from the multi-wavelength SED that Source A is an AGN.  
Existing X-ray coverage of the field reveals no X-ray detection at the position of 
the system \citep[$F_{X}>1.5\times10^{-14}$ erg~s$^{-1}$ cm$^{-2}$ or $L_{X}>2.7\times10^{44}$ 
erg~s$^{-1}$, 2-7 keV observed;][]{ken05} but 
is too shallow to rule out lower lumiosity Seyfert galaxies.  
The typical luminosities of Seyfert galaxies: $L_{Ly\alpha} = 10^{42} - 2\times10^{44}$ erg~s$^{-1}$ and 
$L_{X}$ (0.5-4.5 keV) = $5\times10^{42} - 10^{45}$ erg/s, with $L_{Ly\alpha}/L_{X}$ ratios of $\sim0.1-2$ 
\citep{kri84}.  If we combine 
the $L_{Ly\alpha}/L_{X}$ ratio as an upper limit (since the ratio for an 
extended nebula will be smaller due to the smaller covering fraction of the gas) 
with our measured Lya luminosity ($5.4\times10^{43}$ erg~s$^{-1}$), we should expect 
$L_{X}$ (0.5-4.5 keV)$ > 2.7\times10^{43}-5.4\times10^{44}$ 
erg~s$^{-1}$, which is at or below our current X-ray detection threshold.  

The optical and MIR photometric measurements for Source A are listed in Table~\ref{tab:phot}.  
We measured the broad-band optical fluxes of Source A from NDWFS using 2.1\arcsec\ 
diameter apertures and applied aperture 
corrections of [1.06, 1.40, 1.07] in the \bw, $R$, and $I$ bands.  
The Source A IRAC photometry comes from the Spitzer Deep Wide-Field Survey (SDWFS; 3.5\arcsec\ 
diameter aperture with point source aperture corrections of [1.38,1.38,1.38,1.42]).  
The IRAC colors of Source A lie outside the AGN color-color selection regions of \citet{stern05} and \citet{lac04}, 
and the probability that an X-ray AGN will have these IRAC colors is small \citep{gor08}.  
The IRAC colors of Source A are $[3.6]-[4.5]=0.52\pm0.03$ (Vega) and $[5.8]-[8.0]=0.131\pm0.11$ (Vega) 
\citep[in ``Region B" of Figure 4 of][]{gor08}.  The percentage of X-ray sources down to the XBo\"otes limit 
with IRAC colors in this region is 4\%.  In a small portion 
of the survey with 10 times deeper X-ray coverage, the distribution of sources in IRAC color-color space is similar, i.e., 
there are very few X-ray sources with the IRAC colors of Source A \citep{gor08}.
In contrast, these IRAC colors are consistent with star-forming galaxies at 
$z=1.25-1.75$ \citep{don08}, the redshift range of our source.  
Similarly, the IRAC SED of Source A shows the 1.6$\mu$m bump rather than 
the power-law typical of obscured AGN \citep{alo06}, 
indicating the MIR SED is dominated by stellar emission.  
All of the other sources in the vicinity show similar non-power-law SEDs.

PRG1 is also undetected at longer wavelengths: 
the MIPS 24~$\mu$m non-detection corresponds to an upper limit of 51~$\mu$Jy (1$\sigma$), and 
the Westerbork 20~cm survey non-detection yields a 
5$\sigma$ limit at 3~GHz in the restframe of 
$6.4\times10^{23}$ W~Hz$^{-1}$ \citep{dev02}, well below the realm of high redshift radio galaxies 
\citep[$\sim10^{26}$ W~Hz$^{-1}$;][]{sey07}.

On the other hand, the energetics of the nebula suggest that an AGN at the projected position of Source A 
could explain the observed continuum emission if the AGN is highly obscured to our line-of-sight.  
If we make the assumption that an AGN at Source A is powering the \heii\ emission, 
we can estimate the amount of \bw\ continuum emission we expect from the nebula due to illumination by 
the AGN.  Scaling a standard AGN template \citep{elv94} to match the ionizing photon flux implied by the \heii\ 
(Q(He$^{+})\approx6.2\times10^{53}$~photons~s$^{-1}$), we estimate that the \bw\ luminosity 
from the AGN striking the cloud should be $L_{B_{W},incident}\approx$1.9$\times$10$^{41}$~erg~s$^{-1}$~\AA$^{-1}$.  
This incident AGN continuum flux $-$ the maximum possible contribution from the AGN to the observed continuum emission $-$ 
is roughly the same as the measured \bw\ continuum emission from the nebula 
($L_{B_{W},nebula}\approx$1.1$\times$10$^{41}$~erg~s$^{-1}$ \AA$^{-1}$).  
In reality, the continuum observed within the nebula will arise solely from two-photon (2$\gamma$) 
continuum and scattering of AGN light with no contribution directly from the AGN itself.    
Given the predicted \heii\ equivalent width from our {\tt Cloudy} models of AGN illumination, 
we estimate the expected \bw\ two-photon continuum within the nebula 
to be $L_{B_{W},2\gamma}\approx$2.0$\times$10$^{40}$~erg~s$^{-1}$, 
roughly 18\% of the observed \bw\ continuum of the nebula.  
Electron scattering of AGN light is expected to have a scattering optical depth 
of $\tau=n_{e}l\sigma_{T}$=0.1-1.1, where $n_{e}\approx$1-10~cm$^{-3}$ and $l=56$~kpc is the approximate 
path through the nebula, corresponding to a Thompson scattering continuum of 
$L_{B_{W},Thompson}\lesssim$1.9$\times$10$^{41}$~erg~s$^{-1}$ \AA$^{-1}\times(1-e^{-\tau})$ = 
1.3$\times$10$^{41}$~erg~s$^{-1}$ \AA$^{-1}$, comparable to the observed \bw\ continuum emission.  
Although dust scattering, which is much more efficient than Thompson scattering, may contribute as well, large 
quantities of dust would be difficult to reconcile with the large observed \lya\ and \heii\ luminosities.  
Questions remain, but given the limitations of our discovery data it appears that an AGN at the position of Source A 
that is unobscured along the line-of-sight to the nebula, but completely hidden from our viewpoint, 
could explain the observed \bw\ continuum emission.  
Correcting for the maximal (since we are assuming the projected separation) 
solid angle subtended by the cloud as seen from the Source A (d$\Omega\approx3.1$ steradians), 
we find that the minimum intrinsic AGN luminosity needed to produce this incident flux 
is $L_{B_{W},min(AGN)}\gtrsim$1.9$\times$10$^{41}$~erg~s$^{-1}\times$4$\pi$/d$\Omega$ 
erg~s$^{-1}$ \AA$^{-1}\approx$7.6$\times$10$^{41}$~erg~s$^{-1}\times$ \AA$^{-1}$.  
Source A has a \bw\ luminosity of 1.8$\times$10$^{40}$ erg~s$^{-1}\times$ \AA$^{-1}$, which is over an order of magnitude fainter 
than this firm lower limit.  The AGN would therefore need to be very highly obscured along our 
line-of-sight to match the existing observational constraints.

In conclusion, an AGN at the position of Source A is a plausible source of the hard ionizing radiation.  
However, there is no visible AGN in the vicinity of the nebula, so it would have to be highly obscured along 
our line-of-sight.  An AGN in the vicinity of the nebula must be in a radio-quiet phase and 
so highly obscured that even the observed MIR SED is dominated by light from the host galaxy.  
Deep optical and near-infrared spatially-resolved spectroscopy will be required to resolve this issue.

\subsubsection{Shocks}

The observed ratios are inconsistent with shock ionization in solar metallicity gas, which 
typically produces much higher \ciii/\heii\ and \civ/\ciii\ ratios 
\citep[e.g., $\sim$3-25 and $\sim$1-10 for shock velocities of 500-150~km~s$^{-1}$; ][]{dop96} 
along with strong \nv$\lambda$1240.  
Stronger shock velocities are inconsistent with the narrow 
width of the \heii\ line ($v_{FWHM}\lesssim$500~km~s$^{-1}$).  

\subsubsection{Wolf-Rayet Stars}

Strong \heii\ emission is seen in both of the major classes of Wolf-Rayet stars (WN and WC), 
but it is accompanied by strong \siiv, \nv, and/or other metal emission lines.  
WN stars show strong \siiv$\lambda\lambda$1393,1402 relative to \heii, \civ, and \ciii, 
which we can rule out with our discovery spectra.  
WC stars usually show \civ/\ciii\ ratios greater than 1, medium-strength \siiv, and 
a large number of other metal lines (e.g., Fe), all of which are inconsistent with our observations.  
The spectrum of a typical `W-R galaxy' (a galaxy with spectrum dominated by Wolf-Rayet features) 
effectively averages over hundreds or thousands of W-R stars, 
but none-the-less typically shows \siiv, \civ, \heii, and \nv\ emission with P-Cygni profiles due to the 
effects of strong stellar winds \citep[e.g., ][]{lei96}.  
The composite Lyman Break Galaxy (LBG) spectrum measured at $z\sim3$ also shows \siiv\ and \civ\ 
with P-Cygni profiles from the stellar winds of massive stars, as well as broad \heii\ 
($v_{FWHM}\sim$1500 km~s$^{-1}$), which the authors argue is most likely due to the fast, dense winds of 
Wolf-Rayet stars \citep{shap03}.  In our case, the narrow width of the \heii\ line ($v_{FWHM}\lesssim$500~km s$^{-1}$) and the 
absence of other important Wolf-Rayet features (e.g., \siiv, \nv) rules out the idea that the \heii\ emission 
is coming from Wolf-Rayet stars.  

\subsubsection{Population~II Star Formation}

The existence of diffuse, blue continuum that is cospatial with the \lya\ and \heii\ line emission (Figures~\ref{fig:prg1postage}) 
is suggestive of a distributed source of ionization such as spatially-extended star formation.  
However, the observed \heii/\lya\ line ratio is inconsistent with ionization 
by a standard Population~II SED (e.g., $Z=0.001=1/20 Z_{\sun}$).  The expected Q(He$^{+}$)/Q(H) ratio for a \popii\ model is 
several orders of magnitude lower than observed \citep{sch03}.  
Predictions from our {\tt Cloudy} models indicate that the dearth of hard ionizing photons 
translates into negligible \heii\ emission and \heii/\lya\ ratios that are a factor of 1000 lower 
than observed.  Furthermore, the large observed \heii\ equivalent width can only be produced by very low metallicity 
stellar populations \citep{sch03}.

\subsubsection{Population~III Star Formation}

The only way to explain strong \heii\ emission with ionization from a stellar population 
is to invoke very young ages and very low metallicities.  
The rest-frame equivalent width of \lya\ (EW$\approx$294\AA) is higher than or comparable to what 
is expected from solar metallicity and \popii\ stellar populations \citep{sch03}, but it is by 
no means the largest observed in \lya-emitting galaxy surveys \citep[EW$_{max}\gtrsim$300\AA;][]{daw07}.  
The measured equivalent width of \heii, however, is very large (EW$\approx$37\AA) 
in the context of stellar populations 
and only consistent with the youngest ($\lesssim$2~Myr) and lowest metallicity stars ($\lesssim$10$^{-7}$~Z$_{\sun}$), in 
the absence of mass loss \citep{sch03}.  Here we chose a \popiii\ model with a moderately 
top-heavy IMF from \citet{tum06} that has a peak at 10~$M_{\sun}$ (their case A); a more top-heavy IMF will 
tend to boost the \heii/\lya\ ratio due to the additional hard ionizing photons.  
The ionizing photon flux implied by the \lya\ and \heii\ measurements from PRG1 implies 
a \popiii\ cluster mass of 13-100
$\times$10$^{6}$~M$_{\sun}$ and a \bw\ continuum flux of 
$L_{B_{W},incident}\approx$1.7$\times$10$^{41}$~erg~s$^{-1}$ \AA$^{-1}$, roughly equal to the observed continuum 
($L_{B_{W},nebula}\approx$1.1$\times$10$^{41}$~erg~s$^{-1}$ \AA$^{-1}$).  
In comparing with our {\tt Cloudy} models, however, we find that the observed \heii/\lya\ ratio ($\approx$0.12) is much higher 
than expected from our \popiii\ model ($\approx$0.004).  However, it is important to keep in mind that 
model uncertainties, such as the assumed IMF and the effects of mass loss, could have a large effect 
on the predicted line luminosities.  In addition, the possible underestimation of the \lya\ flux discussed in 
Section 4.1 as well as a density-bounded geometry will tend to boost the observed \heii/\lya\ ratio.  

On the face of it, it would be surprising to find such low metallicity star formation at such a low redshift.  
However, we cannot conclusively rule out the \popiii\ scenario on this basis alone.  
While the mean metallicity of the Universe increases with time, several theoretical models of 
\popiii\ star formation have suggested that significant metallicity inhomogeneities may exist even at 
low redshifts \citep{tor07}.  These models predict that while the metallicity is quickly enriched at the center of 
collapsed structures, low metallicity regions can persist on the periphery over longer timescales, 
allowing \popiii\ star formation to proceed well after the epoch of reionization \citep{tor07,sca03}.  
At $z\approx2.3$ (roughly the redshift midpoint of our \lya\ nebula survey), 
\citet{tor07} predict a \popiii\ SFR density of 1.3$\times10^{-7}~M_{\sun}$~yr$^{-1}$~Mpc$^{-3}$.  
If we were to assume that this \lya+\heii\ nebula is powered 
by very low metallicity star formation, use the \lya\ line emission (extrapolated to the entire nebula 
and corrected for absorption) to estimate a SFR  
\citep[SFR = L$_{Ly\alpha}$ / 1.26$\times10^{42}$ $\approx$ 5.4$\times10^{43}$ / 1.26$\times10^{42}$ erg s$^{-1}$ $\approx$ 42.9 
$M_{\sun}$~yr$^{-1}$;][]{ken98}, 
and ignore any incompleteness of our survey, we would calculate that one nebula of this kind within 
our survey volume (1.3$\times10^{8}$~Mpc$^{3}$), corresponds to a \popiii\ SFR density of 
$\sim3.3\times10^{-7}~M_{\sun}$~yr$^{-1}$~Mpc$^{-3}$.  This very rough estimate based on a single source is of 
the same order of magnitude as the model predictions (within a factor of 3) despite the large theoretical 
uncertainties in the conditions regulating \popiii\ star formation and IGM enrichment at all redshifts.  

In reality, the amount of low metallicity gas and \popiii\ star formation as a function of redshift will depend 
on the efficiency of star formation in different environments and the efficiency 
with which stars pollute their environment, i.e., the feedback efficiency.  
A more realistic treatment of galactic winds in cosmological simulations 
\citep[e.g., momentum-driven winds;][]{ope06} is shown to suppress the 
metallicity in all gas phases relative to the constant wind scenario similar to that used in \citet{tor07}, which 
may in turn increase the fraction of \popiii\ star formation at any given redshift.  
On the other hand, only a few stellar generations would be required to produce 
the mass of carbon in the nebula.  
Assuming the derived ionized gas mass ($\sim8\times10^{8}$ M$_{\odot}$), the simulated 
yields for \popiii\ supernovae \citep[$\sim$0.2-1 M$_{\odot}$ C per 20-40 M$_{\odot}$ 
supernova progenitor;][]{tom07} and instantaneous mixing, enriching the cloud to 
$z\approx10^{-3} Z_{\odot}$ would only require of order 750 supernovae 
($2-4\times10^{4}$ M$_{\odot}$).  This is several orders of magnitude lower than the 
fraction of the cluster mass implied by the \lya\ and \heii\ luminosities that is in stars 
with $M>20M_{\odot}$ for the assumed top-heavy IMF: $f(M>20M_{\odot})=0.26$ or $3-27\times10^{6} M_{\odot}$.  
Thus it is likely the nebula would be polluted to the observed levels over 
a relatively short timescale.  

Large uncertainties remain in our understanding of \popiii\ star formation.  
The expected fraction of \popiii\ star formation at a give epoch is unclear, 
feedback efficiency of these first stars is largely unconstrained, 
and there are a wide range of possible 
\lya/\heii\ ratios, depending on stellar mass loss, star formation 
history, and the complicated effects of \lya\ radiative transfer.  
Despite the strong \lya+\heii\ signature in this source, 
we cannot conclusively confirm or rule out \popiii\ star formation as a source of ionization 
with the current observations.   

\subsubsection{Gravitational Cooling Radiation}

A number of theoretical papers have addressed the issue of gravitational cooling 
radiation, i.e., the cooling of gas as it collapses within the dark matter potential and heats 
to $T\approx10^{4}$~K.  Metal-line cooling is possible for gas with  $T<10^4$~K, but it 
is $\sim$1000 times less efficient than cooling via \lya\ emission, unless $Z>0.1Z_{\sun}$ \citep{hai00}.  
Thus for low metallicity gas, strong \lya\ emission is expected to dominate the cooling \citep{far01}.  
\heii\ emission may be important as well, but its contribution relative to \lya\ 
is unclear due to uncertainties in how much \lya\ will escape the system.  
From a semi-analytic calculation assuming monolithic collapse, \citet{hai00} 
suggested that a cloud that is sufficiently metal-poor will radiate 10\% of the energy via 
\heii$\lambda$304\AA; this corresponds to a \heii$\lambda$1640/\lya\ of $\sim$0.01, 
where we have followed \citet{yang06} and adopted a ratio of \heii$\lambda$1640/\heii$\lambda$304$\approx$0.10, 
i.e., the case B values of \citet{stor95} extrapolated to the low density limit.  
Using Parallel TreeSPH simulations, \citet{yang06} predicted ratios 
of \heii$\lambda$1640/\lya$\lesssim0.10-0.001$ for gravitationally cooling clouds, depending 
on the degree to which self-shielding of the gas reduces the \lya\ flux.  This range approaches the 
observed ratio for PRG1, but due to subsequent corrections to the treatment of 
star-forming gas in more recent simulations 
the predictions from \citet{yang06} are likely overestimates (Y. Yang 2008, private communication).  
Thus, the observed \heii/\lya\ ratio appears to be higher than predicted for 
gravitational cooling radiation, but again the uncertainties in both the theoretical predictions 
and in our \lya\ measurement make it difficult to draw firm conclusions.  

Some of the theoretical predictions for gravitationally cooling clouds are consistent with our observations.  
The observed \heii\ line width ($\sigma_{v}<250$~km~s$^{-1}$) is more 
consistent with gravitational cooling radiation \citep[$\sigma_{v}<400$~km~s$^{-1}$; e.g.,][]{yang06} 
than with outflows.  In addition, theoretical simulations of the redshift evolution 
predict a peak in the number density of gravitational cooling \lya\ nebulae at $z\approx2$, 
consistent with our discovery of a \lya+\heii\ nebula at $z\approx1.7$.  
While \citet{dijk07} predicts rest-frame \lya\ equivalent widths of $>$1000\AA\ 
for gravitationally cooling clouds, he notes that the observed rest-frame equivalent width is 
likely to be reduced by a factor of 5-10 due to scattering of \lya\ photons in the intergalactic medium, 
more in line with our measurements.  

However, a number of other theoretical predictions of gravitational cooling radiation do not fit our observations.  
First, the simulations do not reproduce the relative sizes of the observed \lya\ and \heii\ regions.  
\citet{far01} could not reproduce the sizes of the largest observed \lya\ nebulae unless they invoked 
resonant scattering of the \lya\ emission.  In that case, the \heii\ emission should be 
more centrally-concentrated than the \lya.  Similarly, \citet{yang06} suggested that \heii\ regions 
would likely be too small to resolve with current observational facilities.  
In contrast, in our \lya+\heii\ nebula the \heii\ region appears to be comparable in size 
to the \lya\ region ($\approx$45~kpc).  
The theoretical papers also predict that \lya\ nebulae will only be present 
as a halo around a massive galaxy \citep{far01,fur05}.  While there are a few faint sources 
around the edge of our nebula, and there may be unresolved low surface brightness clumps within the cloud, 
the multi-wavelength imaging shows no evidence for a central massive galaxy in this system.  

Given the uncertainties in the theoretical predictions and the limits of our discovery data 
it is difficult to assess the applicability of the gravitational cooling model.  
The weight of the current evidence disfavors gravitational cooling as 
the sole explanation for the line emission, but it is certainly possible that the 
nebula is powered by multiple processes, with gravitational cooling radiation playing a larger 
role on the outskirts and photoionization from stars or AGN dominating the ionization of the 
central regions.
 
\section{Implications}

Strong \lya+\heii\ in the absence of strong metal lines has been championed as a unique 
observational signature of primordial galaxy formation (e.g., \popiii\ star formation 
or gravitationally cooling clouds), but the discovery of this \lya+\heii\ nebula suggests that 
the situation can be much more complex in actual astrophysical sources.  
Occam's razor suggests that the most likely power source is an AGN at the position of Source A.  
An analysis of the existing data shows no obvious evidence of a powerful AGN in the vicinity, 
so to explain the ionization of the nebula, an AGN would need to be highly obscured 
even in the mid-infrared.  
The line ratios rule out ionization by shocks, Wolf-Rayet stars, and \popii\ star formation.  
\popiii\ star formation would provide the necessary 
hard ionizing continuum to explain the observed line ratios along with a natural explanation for 
the spatially-extended continuum emission, but this scenario appears unlikely given 
the low redshift.  Despite the compelling \lya+\heii\ signature, the complexity of this source and 
the large uncertainties in theoretical predictions preclude a more definitive conclusion.
The contribution from gravitational cooling radiation is similarly unclear, although 
the morphology of the nebula (with no central compact source) 
and relative sizes of the \lya\ and \heii\ emitting regions disfavor this 
hypothesis as a dominant mechanism.  

One of the most important implications of this discovery is that it demonstrates we must 
be careful when using strong \lya+\heii\ emission as a tracer of \popiii\ star formation.   
Surveys looking specifically for this \lya+\heii\ signature are ongoing.  
While the low redshift of PRG1 allows for the detailed multi-wavelength follow-up necessary in order 
to understand the power source and the metallicity of the gas, the same cannot be said for ongoing \lya+\heii\ 
emitter surveys at higher redshift that will lack longer wavelength coverage and be sensitivity-challenged.  
It is extremely important to note that if this \lya+\heii\ had been discovered at high redshift, it 
would have been easy to jump prematurely to the \popiii\ conclusion.  
The discovery of a \lya+\heii\ nebula at $z\approx1.67$ 
therefore provides an ideal opportunity to evaluate the extent to which strong 
\lya+\heii\ emission can be used as a unique tracer of \popiii\ star formation 
and underscores the importance of using care when interpreting a strong \lya+\heii\ signature 
in the absence of more extensive multi-wavelength data.

\section{Summary}

We have discovered a \lya\ nebula at $z\approx1.67$ (the lowest redshift \lya\ nebula 
known) that shows strong, diffuse \heii\ emission and weak/negligible \ciii\ and \civ\ emission.  
From the line ratios we derive evidence that this nebula may contain 
low metallicity ($Z<10^{-2}-10^{-3} Z_{\sun}$) gas, depending on the unknown ionization parameter, 
that is being illuminated by a hard ionizing continuum, either due an AGN or a very low metallicity 
stellar population (\popiii), by gravitational cooling, or some combination thereof.  
The softer continua of Population~I and II stars can be conclusively ruled out along 
with ionization by shocks and Wolf-Rayet stars.  No obvious, unobscured, powerful AGN is seen 
in the system; thus if an AGN is responsible, it must be highly obscured along our line-of-sight.  
Despite the strong \lya+\heii\ signature, our detailed analysis of the discovery data 
shows that dedicated follow-up observations will be required in order to draw firm conclusions 
about the dominant source of ionization for the nebula and better constrain the metallicity.  
This is the first time that the predicted observational signature of very low metallicity (\popiii) star formation $-$ 
strong \lya\ and \heii\ in the absence of strong metal lines $-$ has been seen in a spatially-extended source; 
however, the complex nature of the nebula and the fact that such complexity becomes 
increasingly difficult to discern with redshift suggest that studies at high redshift will need to 
use caution when interpreting future \lya+\heii\ discoveries.

\acknowledgments

We are grateful to J. Tumlinson for providing the \popiii\ evolving spectra, to E. Le Floc'h and the IRS GTO and 
MIPS GTO teams for providing the MIPS data, to Gary Ferland for useful advice on the photoionization modeling, 
and to K. Finlator for observing assistance and many helpful discussions.  D. Stern, M. Ashby, M. Brodwin, and the rest 
of the SDWFS team are thanked for access to the most current Spitzer IRAC imaging of the Bo\"otes field.  
We also thank the anonymous referee for suggestions that improved the clarity of this paper.  
We would like to acknowledge the expert 
assistance of the staff of the MMT Observatory (especially John McAfee, Ale Milone, and G. Grant Williams).  
M. P. was supported by an NSF Graduate Research Fellowship and a P.E.O Scholar Award.  
This research builds on data from the NOAO Deep Wide-Field Survey (NDWFS) as distributed by the 
NOAO Science Archive.  NOAO is operated by the Association of Universities for 
Research in Astronomy (AURA), Inc. under a cooperative agreement with the National 
Science Foundation.  Some of the data presented in this paper were obtained from the 
Multimission Archive at the Space Telescope Science Institute (MAST). STScI is operated 
by the Association of Universities for Research in Astronomy, Inc., under NASA contract 
NAS5-26555. Support for MAST for non-HST data is provided by the NASA Office of Space 
Science via grant NAG5-7584 and by other grants and contracts.

\begin{deluxetable}{cccccccc}
\tabletypesize{\scriptsize}
\tablecaption{PRG1 Spectroscopic Measurements}
\tablewidth{0pt}
%\rotate
\tablehead{
 & \colhead{Ly$\alpha\lambda$1216} & \colhead{N\textsc{v}$\lambda$1240} & \colhead{Si\textsc{iv}$\lambda$1398} & \colhead{C\textsc{iv}$\lambda$1549} & \colhead{He\textsc{ii}$\lambda$1640} & \colhead{C\textsc{iii}]$\lambda$1909} & \colhead{Ne\textsc{iv}$\lambda$2424} \\
 }
\startdata
\cutinhead{Night 1 - UT 2008 June 08}
          Flux\tablenotemark{a} [$10^{-17}$ erg s$^{-1}$ cm$^{-2}$] &     49.9     $\pm$     5.7     & $<$     6.4  \tablenotemark{b} & $<$     1.8  \tablenotemark{b} & $<$     1.4     \tablenotemark{b} &      6.2     $\pm$     1.7     & $<$     1.2  \tablenotemark{b} & $<$     1.4     \tablenotemark{b} \\
                                                  EW$_{rest}$ [\AA] &    294.1     $\pm$    39.4      & - & - & - &     36.8     $\pm$    10.1                & - & -\\
                                              $\lambda_{obs}$ [\AA] &  3250.07     $\pm$    0.56      & - & - & - &  4383.07     $\pm$    1.25                & - & -\\
                                                           Redshift &   1.6735     $\pm$  0.0005      & - & - & - &   1.6714     $\pm$  0.0008                & - & -\\
                                                  FWHM$_{obs}$ [\AA]&    10.20     $\pm$    0.82      & - & - &- &      8.41     $\pm$    3.23                & - & -\\
                                                  FWHM [km s$^{-1}$]&    941.5     $\pm$    75.5      & - & - &- &     575.5     $\pm$   221.1                & - & -\\

\cutinhead{Night 2 - UT 2008 June 09}
          Flux\tablenotemark{a} [$10^{-17}$ erg s$^{-1}$ cm$^{-2}$] &     43.6     $\pm$     4.0     & $<$     5.5  \tablenotemark{b} & $<$     1.1     \tablenotemark{b} &      2.1     $\pm$     1.1        &      5.7     $\pm$     0.9        &      4.7     $\pm$     0.8        &      2.8     $\pm$     1.0                       \\
                                              EW$_{rest}$ [\AA]     &    257.1     $\pm$    29.4          & - & - &     12.4     $\pm$     6.5        &     33.9     $\pm$     6.1        &     28.0     $\pm$     5.3        &     29.9     $\pm$    11.3                  \\
                                              $\lambda_{obs}$ [\AA] &  3249.59     $\pm$    0.38          & - & - &  4142.30     $\pm$    1.46        &  4381.92     $\pm$    0.76        &  5095.26     $\pm$    0.10        &  6476.81     $\pm$    1.83                  \\
                                              Redshift              &   1.6731     $\pm$  0.0003          & - & - &   1.6724     $\pm$  0.0009        &   1.6707     $\pm$  0.0005        &   1.6695     $\pm$  0.0001        &   1.6720     $\pm$  0.0008                  \\
                                      FWHM$_{obs}$ [\AA]            &     9.75     $\pm$    0.71    & - & - &  -  &     6.36     $\pm$    0.64        &    23.28     $\pm$    1.94             &  - \\
                                              FWHM [km s$^{-1}$]    &    900.4     $\pm$    65.1     & - & - &  - &    435.5     $\pm$    44.1        &   1370.7     $\pm$   114.5    & - \\
\enddata
\tablenotetext{a}{Flux measured in a 1.5$\times$5.0\arcsec\ aperture.  No correction has been applied for Ly$\alpha$ absorption.}
\tablenotetext{b}{Quoted upper limits are 1$\sigma$ values.}
\label{tab:fluxkin}
\end{deluxetable}

\begin{deluxetable}{ccc}
\tabletypesize{\scriptsize}
\tablecaption{PRG1 Photometric Measurements}
\tablewidth{0pt}
\tablehead{
 & \colhead{Nebula\tablenotemark{a,b}} & \colhead{Source A\tablenotemark{c}} \\
 & \colhead{$10^{-30}$ erg s$^{-1}$ cm$^{-2}$ Hz$^{-1}$} & \colhead{$10^{-30}$ erg s$^{-1}$ cm$^{-2}$ Hz$^{-1}$} \\
 }
\startdata
         $B_{W}$ &    3.58 $\pm$  0.24    &   2.07 $\pm$  0.14                            \\
            $R$  &    4.03 $\pm$  0.49    &   4.60 $\pm$  0.33                            \\
            $I$  &    6.62 $\pm$  0.88    &   8.35 $\pm$  0.47                            \\
3.6$\mu$m  &  -  &  466.56 $\pm$ 34.13                            \\
4.5$\mu$m  &  -  &  488.70 $\pm$ 32.40                            \\
5.8$\mu$m  &  -  &  329.67 $\pm$ 31.40                            \\
8.0$\mu$m  &  -  &  194.13 $\pm$ 25.60                            \\
\enddata
\tablenotetext{a}{Flux measured within the same 1.5$\times$5.0\arcsec\ aperture as the spectroscopic measurements.}
\tablenotetext{b}{The contribution of line emission for Night 1 (Night 2): \heii\ contributes 8\% (7\%) and \civ\ $<2$\% (3\%) of the \bw\ emission, and \neiv\ contributes $<2$\% (6.5\%) of the $R$-band emission.  We see no contribution from line emission out to the middle of the $I$-band (8300\AA, the extent of our spectroscopic coverage).}
\tablenotetext{c}{Optical \bw, $R$, and $I$ fluxes were measured within 2.1\arcsec\ diameter apertures.  Mid-infrared fluxes (3.6, 4.5, 5.8, 8.0 $\mu$m) were measured within 3.5\arcsec\ diameter apertures.  Aperture corrections are discussed in the text.}
\label{tab:phot}
\end{deluxetable}

\begin{deluxetable}{cccccc}
\tabletypesize{\scriptsize}
\tablecaption{Emission Line Ratio Comparison}
\tablewidth{0pt}
\tablehead{
 & \colhead{Ly$\alpha$/He\textsc{ii}\tablenotemark{a}} & \colhead{C\textsc{iv}/He\textsc{ii}} & \colhead{C\textsc{iii}]/He\textsc{ii}} & \colhead{C\textsc{iv}/C\textsc{iii}]} & \colhead{Reference}\\
 }
\startdata
 Radio Galaxy Halos (Composite)   & 11.7 & 1.75 & 0.73 & 2.4 & Humphrey et al. 2008\tablenotemark{b} \\
            Radio Galaxy Halos (Mean) &   9.80 $\pm$  5.69  &   1.59 $\pm$  0.56 &   1.06 $\pm$  1.05 &   2.22 $\pm$  1.17      & Humphrey et al. 2008\\
     Stacked Lya Blobs at $z\approx3$ &  11.11 $\pm$  9.88 &                 -  &                 -  &                 -          & Saito et al. 2008\\
     Lya Blob at $z\approx2.7$        &   7.62 $\pm$  0.08  &   1.02 $\pm$  0.01 &   0.12 $\pm$  0.02 &   8.34 $\pm$  1.67           & Dey et al. 2005\\
     PRG1 Night 1                     &   8.00 $\pm$  2.32                    & $<$  0.22  \tablenotemark{c} & $<$  0.19    \tablenotemark{c} & - & This study\\
     PRG1 Night 2                     &   7.59 $\pm$  1.43  &   0.36 $\pm$  0.20 &   0.83 $\pm$  0.20 &   0.44 $\pm$  0.24                & This study\\
\enddata
\tablenotetext{a}{No correction has been applied for Ly$\alpha$ absorption.}
\tablenotetext{b}{Errors on line ratios from composite spectrum not given.}
\tablenotetext{c}{1$\sigma$ upper limits.}
\label{tab:lineratio}
\end{deluxetable}


\begin{thebibliography}{}
\bibitem[Alonso-Herrero et al.(2006)]{alo06} Alonso-Herrero, A., et al.\ 2006, \apj, 640, 167 
\bibitem[Bromm et al.(2001)]{bro01} Bromm, V., Kudritzki, R.~P., \& Loeb, A.\ 2001, \apj, 552, 464 
\bibitem[Dawson et al.(2004)]{daw04} Dawson, S., et al.\ 2004, \apj, 617, 707 
\bibitem[Dawson et al.(2007)]{daw07} Dawson, S., Rhoads, J.~E., Malhotra, S., Stern, D., Wang, J., Dey, A., Spinrad, H., \& Jannuzi, B.~T.\ 2007, \apj, 671, 1227 
\bibitem[Dey et al.(1997)]{dey97} Dey, A., van Breugel, W., Vacca, W.~D., \& Antonucci, R.\ 1997, \apj, 490, 698 
\bibitem[Dey et al.(2005)]{dey05} Dey, A., et al.\ 2005, \apj, 629, 654 
\bibitem[Dijkstra(2007)]{dijk07} Dijkstra, M.\ 2007, ArXiv e-prints, 711, arXiv:0711.2698 
\bibitem[di Serego Alighieri et al.(2008)]{ali08} di Serego Alighieri, S., Kurk, J., Ciardi, B., Cimatti, A., Daddi, E., \& Ferrara, A.\ 2008, ArXiv e-prints, 807, arXiv:0807.4634 
\bibitem[Donley et al.(2008)]{don08} Donley, J.~L., Rieke, G.~H., Perez-Gonzalez, P.~G., \& Barro, G.\ 2008, ArXiv e-prints, 806, arXiv:0806.4610 
\bibitem[Dopita \& Sutherland(1996)]{dop96} Dopita, M.~A., \& Sutherland, R.~S.\ 1996, \apjs, 102, 161 
\bibitem[de Vries et al.(2002)]{dev02} de Vries, W.~H., Morganti, R., R{\"o}ttgering, H.~J.~A., Vermeulen, R., van Breugel, W., Rengelink, R., \& Jarvis, M.~J.\ 2002, \aj, 123, 1784 
\bibitem[Elvis et al.(1994)]{elv94} Elvis, M., et al.\ 1994, \apjs, 95, 1 
\bibitem[Ezer \& Cameron(1971)]{ez71} Ezer, D., \& Cameron, A.~G.~W.\ 1971, \apss, 14, 399 
\bibitem[Fardal et al.(2001)]{far01} Fardal, M.~A., Katz, N., Gardner, J.~P., Hernquist, L., Weinberg, D.~H., \& Dav{\'e}, R.\ 2001, \apj, 562, 605 
\bibitem[Ferland \& Osterbrock(1986)]{fer86} Ferland, G.~J., \& Osterbrock, D.~E.\ 1986, \apj, 300, 658
\bibitem[Ferland et al.(1998)]{fer98} Ferland, G.~J., Korista, K.~T., Verner, D.~A., Ferguson, J.~W., Kingdon, J.~B., \& Verner, E.~M.\ 1998, \pasp, 110, 761 
\bibitem[Finkelstein et al.(2008)]{fink08} Finkelstein, S.~L., Rhoads, J.~E., Malhotra, S., Grogin, N., \& Wang, J.\ 2008, \apj, 678, 655 
\bibitem[Furlanetto et al.(2005)]{fur05} Furlanetto, S.~R., Schaye, J., Springel, V., \& Hernquist, L.\ 2005, \apj, 622, 7 
\bibitem[Geach et al.(2007)]{gea07} Geach, J.~E., Smail, I., Chapman, S.~C., Alexander, D.~M., Blain, A.~W., Stott, J.~P., \& Ivison, R.~J.\ 2007, \apjl, 655, L9 
\bibitem[Gorjian et al.(2008)]{gor08} Gorjian, V., et al.\ 2008, \apj, 679, 1040 
\bibitem[Haiman et al.(2000)]{hai00} Haiman, Z., Spaans, M., \& Quataert, E.\ 2000, \apjl, 537, L5 
\bibitem[Heckman et al.(1989)]{heck89} Heckman, T.~M., Baum, S.~A., van Breugel, W.~J.~M., \& McCarthy, P.\ 1989, \apj, 338, 48 
\bibitem[Humphrey et al.(2008)]{hum08} Humphrey, A., Villar-Mart{\'{\i}}n, M., Vernet, J., Fosbury, R., di Serego Alighieri, S., \& Binette, L.\ 2008, \mnras, 383, 11 
\bibitem[Jannuzi \& Dey(1999)]{jan99} Jannuzi, B.~T., \& Dey, A.\ 1999, in ASP Conf. Ser. 191, Photometric Redshifts and the Detection of High Redshift Galaxies, ed. R. Weymann et al. (San Francisco: ASP), 111 
\bibitem[Kennicutt(1998)]{ken98} Kennicutt, R.~C., Jr.\ 1998, \apj, 498, 541 
\bibitem[Kenter et al.(2005)]{ken05} Kenter, A., et al.\ 2005, \apjs, 161, 9 
\bibitem[Kriss(1984)]{kri84} Kriss, G.~A.\ 1984, \apj, 277, 495 
\bibitem[Lacy et al.(2004)]{lac04} Lacy, M., et al.\ 2004, \apjs, 154, 166 
\bibitem[Leitherer et al.(1996)]{lei96} Leitherer, C., Vacca, W.~D., Conti, P.~S., Filippenko, A.~V., Robert, C., \& Sargent, W.~L.~W.\ 1996, \apj, 465, 717 
\bibitem[Lynch \& Charlton(2007)]{lyn07} Lynch, R.~S., \& Charlton, J.~C.\ 2007, \apj, 666, 64
\bibitem[Malhotra \& Rhoads(2002)]{mal02} Malhotra, S., \& Rhoads, J.~E.\ 2002, \apjl, 565, L71 
\bibitem[Martin et al.(2005)]{mar05} Martin, D.~C., et al.\ 2005, \apjl, 619, L1 
\bibitem[Mathews \& Ferland(1987)]{mat87} Mathews, W.~G., \& Ferland, G.~J.\ 1987, \apj, 323, 456 
\bibitem[Matsuda et al.(2007)]{mat07} Matsuda, Y., Iono, D., Ohta, K., Yamada, T., Kawabe, R., Hayashino, T., Peck, A.~B., \& Petitpas, G.~R.\ 2007, \apj, 667, 667 
\bibitem[Maxfield et al.(2002)]{max02} Maxfield, L., Spinrad, H., Stern, D., Dey, A., \& Dickinson, M.\ 2002, \aj, 123, 2321 
\bibitem[McCarthy et al.(1987)]{mcc87} McCarthy, P.~J., Spinrad, H., Djorgovski, S., Strauss, M.~A., van Breugel, W., \& Liebert, J.\ 1987, \apjl, 319, L39 
\bibitem[Nagao et al.(2005)]{nag05} Nagao, T., Motohara, K., Maiolino, R., Marconi, A., Taniguchi, Y., Aoki, K., Ajiki, M., \& Shioya, Y.\ 2005, \apjl, 631, L5 
\bibitem[Nagao et al.(2008)]{nag08} Nagao, T., et al.\ 2008, \apj, 680, 100 
\bibitem[Nilsson et al.(2006)]{nil06} Nilsson, K.~K., Fynbo, J.~P.~U., M{\o}ller, P., Sommer-Larsen, J., \& Ledoux, C.\ 2006, \aap, 452, L23 
\bibitem[Oppenheimer \& Dav{\'e}(2006)]{ope06} Oppenheimer, B.~D., \& Dav{\'e}, R.\ 2006, \mnras, 373, 1265
\bibitem[Osterbrock(1989)]{ost89} Osterbrock, D.~E.\ 1989, Research supported by the University of California, John Simon Guggenheim Memorial Foundation, University of Minnesota, et al.~Mill Valley, CA, University Science Books, 1989, 422 p.
\bibitem[Ouchi et al.(2008)]{ouch08} Ouchi, M., et al.\ 2008, \apjs, 176, 301 
\bibitem[Pettini et al.(2008)]{pet08} Pettini, M., Zych, B.~J., Steidel, C.~C., \& Chaffee, F.~H.\ 2008, \mnras, 385, 2011 
\bibitem[Reuland et al.(2003)]{reu03} Reuland, M., et al.\ 2003, \apj, 592, 755 
\bibitem[Reuland et al.(2007)]{reu07} Reuland, M., et al.\ 2007, \aj, 133, 2607 
\bibitem[Rhoads et al.(2003)]{rho03} Rhoads, J.~E., et al.\ 2003, \aj, 125, 1006 
\bibitem[Saito et al.(2008)]{sai08} Saito, T., Shimasaku, K., Okamura, S., Ouchi, M., Akiyama, M., Yoshida, M., \& Ueda, Y.\ 2008, \apj, 675, 1076 
\bibitem[Scannapieco et al.(2003)]{sca03} Scannapieco, E., Schneider, R., \& Ferrara, A.\ 2003, \apj, 589, 35 
\bibitem[Schaerer(2002)]{sch02} Schaerer, D.\ 2002, \aap, 382, 28 
\bibitem[Schaerer(2003)]{sch03} Schaerer, D.\ 2003, \aap, 397, 527 
\bibitem[Schaerer(2008)]{sch08} Schaerer, D.\ 2008, arXiv:0809.1988 
\bibitem[Seymour et al.(2007)]{sey07} Seymour, N., et al.\ 2007, \apjs, 171, 353 
\bibitem[Shapley et al.(2003)]{shap03} Shapley, A.~E., Steidel, C.~C., Pettini, M., \& Adelberger, K.~L.\ 2003, \apj, 588, 65 
\bibitem[Smith \& Jarvis(2007)]{smi07} Smith, D.~J.~B., \& Jarvis, M.~J.\ 2007, \mnras, 378, L49 
\bibitem[Smith et al.(2008)]{smi08} Smith, D.~J.~B., Jarvis, M.~J., Lacy, M., \& Mart{\'{\i}}nez-Sansigre, A.\ 2008, \mnras, 870 
\bibitem[Songaila(2001)]{son01} Songaila, A.\ 2001, \apjl, 561, L153 
\bibitem[Spitzer(1978)]{spi78} Spitzer, L.\ 1978, New York Wiley-Interscience, 1978.~333 p.
\bibitem[Stern et al.(2005)]{stern05} Stern, D., et al.\ 2005, \apj, 631, 163 
\bibitem[Storey \& Hummer(1995)]{stor95} Storey, P.~J., \& Hummer, D.~G.\ 1995, \mnras, 272, 41
\bibitem[Tominaga et al.(2007)]{tom07} Tominaga, N., Umeda, H., \& Nomoto, K.\ 2007, \apj, 660, 516 
\bibitem[Tornatore et al.(2007)]{tor07} Tornatore, L., Borgani, S., Dolag, K., \& Matteucci, F.\ 2007, \mnras, 382, 1050 
\bibitem[Tumlinson et al.(2001)]{tum01} Tumlinson, J., Giroux, M.~L., \& Shull, J.~M.\ 2001, \apjl, 550, L1
\bibitem[Tumlinson et al.(2006)]{tum06} Tumlinson, J. et al. 2006, ApJ, 641, 1
\bibitem[van Ojik et al.(1996)]{vano96} van Ojik, R., Roettgering, H.~J.~A., Carilli, C.~L., Miley, G.~K., Bremer, M.~N., \& Macchetto, F.\ 1996, \aap, 313, 25 
\bibitem[Villar-Mart{\'{\i}}n et al.(2003)]{vil03} Villar-Mart{\'{\i}}n, M., Vernet, J., di Serego Alighieri, S., Fosbury, R., Humphrey, A., Pentericci, L., \& Cohen, M.\ 2003, New Astronomy Review, 47, 291 
\bibitem[Yang et al.(2006)]{yang06} Yang, Y., Zabludoff, A.~I., Dav{\'e}, R., Eisenstein, D.~J., Pinto, P.~A., Katz, N., Weinberg, D.~H., \& Barton, E.~J.\ 2006, \apj, 640, 539 
\end{thebibliography}
\end{document}